\documentclass[prd,aps,twocolumn,floats,floatfix,float,nofootinbib] {revtex4-1}
\usepackage[utf8]{inputenc}
\usepackage{graphicx}
\usepackage{amsmath,amssymb}
\usepackage{amsfonts}
\usepackage{physics}
\usepackage{xspace}
\usepackage{bm}
\usepackage[normalem]{ulem}
\usepackage{mathrsfs}
\usepackage[dvipsnames]{xcolor}
\usepackage[unicode]{hyperref}
\hypersetup{colorlinks=true, citecolor=MidnightBlue,
	linkcolor=CornflowerBlue, urlcolor=Blue, linktocpage=true}
\usepackage[all]{hypcap}
\usepackage{multirow}
\usepackage[format=plain,justification=RaggedRight,labelsep=endash,font=small,labelfont=sc]{caption}
\usepackage[export]{adjustbox}

\usepackage{caption}
\usepackage{subcaption}

\definecolor {darkgreen}{rgb}{0.2,0.7,0.2}

\definecolor {dark}{rgb}{0.43,0.5,0.5}

\newcommand{\eq}{\begin{equation}}
\newcommand{\be}{\begin{equation}}
\newcommand{\eeq}{\end{equation}}
\newcommand{\ee}{\end{equation}}
\newcommand{\de}{\partial}

\newcommand{\gcc}{\mbox{${\rm g}/{\rm cm}^3$}}

\begin{document}

\title{Kinetic screening in nonlinear stellar oscillations and gravitational collapse}
 \author{Miguel Bezares}
 \affiliation{SISSA, Via Bonomea 265, 34136 Trieste, Italy and INFN Sezione di Trieste}
 \affiliation{IFPU - Institute for Fundamental Physics of the Universe, Via Beirut 2, 34014 Trieste, Italy}
  \author{Lotte ter Haar}
 \affiliation{SISSA, Via Bonomea 265, 34136 Trieste, Italy and INFN Sezione di Trieste}
 \affiliation{IFPU - Institute for Fundamental Physics of the Universe, Via Beirut 2, 34014 Trieste, Italy}
 \author{Marco Crisostomi}
 \affiliation{SISSA, Via Bonomea 265, 34136 Trieste, Italy and INFN Sezione di Trieste}
 \affiliation{IFPU - Institute for Fundamental Physics of the Universe, Via Beirut 2, 34014 Trieste, Italy}
  \author{Enrico Barausse}
 \affiliation{SISSA, Via Bonomea 265, 34136 Trieste, Italy and INFN Sezione di Trieste}
 \affiliation{IFPU - Institute for Fundamental Physics of the Universe, Via Beirut 2, 34014 Trieste, Italy}
 \author{Carlos Palenzuela}
\affiliation{Departament  de  F\'{\i}sica,  Universitat  de  les  Illes  Balears  and  Institut  d'Estudis Espacials  de  Catalunya,  Palma  de  Mallorca,  Baleares  E-07122,  Spain}
\affiliation{Institut Aplicacions Computationals (IAC3),  Universitat  de  les  Illes  Balears,  Palma  de  Mallorca,  Baleares  E-07122,  Spain}

\begin{abstract}
We consider $k$-essence, a scalar-tensor theory with first-order derivative self-interactions
that  can screen local scales from scalar fifth forces, while allowing for sizeable deviations from General Relativity on cosmological scales. We construct fully nonlinear static stellar solutions that show the presence of this screening mechanism, and we use them as initial data for simulations
of stellar oscillations and gravitational collapse in spherical symmetry.
We find that  for $k$-essence theories of relevance for cosmology, the screening mechanism works in the case of stellar oscillation and suppresses the monopole scalar emission to undetectable levels. In collapsing stars, we find that the Cauchy problem, although locally well posed, can lead to diverging characteristic speeds for the scalar field. By introducing a ``fixing equation'' in the spirit of J. Cayuso, N. Ortiz, and L. Lehner [Phys. Rev. D 96, 084043 (2017)], inspired in turn by dissipative relativistic hydrodynamics, 
we manage to evolve collapsing neutron stars past the divergence of the characteristic speeds.
We show that, in these systems, the screening mechanism is less efficient
than for oscillating and static stars, because the collapsing star must shed away all of its scalar hair before forming a black hole. 
For  
$k$-essence theories of relevance for cosmology, the characteristic frequency of the
resulting scalar monopole signal is too low  for terrestrial detectors, but
 we conjecture that space-borne interferometers such as LISA might 
detect it if a supernova explodes in the Galaxy.
\end{abstract}
\date{\today \hspace{0.2truecm}}

\maketitle
\flushbottom

\section{Introduction}
General Relativity (GR) has been tested extensively on local scales, e.g. in
 the solar system~\cite{Will:1993hxu,Will:2014kxa}, in binary pulsars~\cite{Damour:1991rd,Kramer:2006nb,Freire:2012mg} 
 and with the Advanced  LIGO/Virgo observations of black-hole and neutron-star binaries~\cite{TheLIGOScientific:2016src,Abbott:2018lct,LIGOScientific:2019fpa,Abbott:2020jks}.
 However, on larger (cosmological) scales, the putative existence of a
 dark sector may hint at a  breakdown of GR in the infrared (see e.g. \cite{Clifton:2011jh} for a review). The obvious difficulty
 of explaining (at least partially) the dark sector as a modification of GR lies
  precisely in the excellent agreement between GR and local observables.
 Therefore, theories that attempt to produce sizeable modifications of the GR phenomenology on
 cosmological scales
 must possess a built-in mechanism screening local scales from large non-GR effects \cite{Clifton:2011jh}.
 
 Among the simplest and most popular theories extending GR are scalar-tensor (ST) theories, where the gravitational interaction is mediated not only by a massless spin-2 field, but also by an additional gravitational scalar. ST theories were first introduced by Fierz~\cite{Fierz:1956zz}, Jordan~\cite{Jordan:1959eg}, Brans and Dicke~\cite{Brans:1961sx} (henceforth FJBD), who proposed the action
\begin{equation}
\label{Jframe_action}
S=\!\!\int d^4 \tilde x\frac{M_\mathrm{Pl}^2 {\sqrt{-\tilde{g}}}}{2}\left[\Phi \tilde{R}-\frac{\omega}{\Phi} \tilde\partial_\mu \Phi \tilde\partial^\mu \Phi
%-V(\phi)
\right]+ S_m[\tilde{g}_{\mu\nu},\Psi_m]\,,
\end{equation}
where $M_\mathrm{Pl}=(8\pi G)^{-1/2}$ is the Planck mass, 
$\tilde{R}$ and $\tilde{g}$ are the Ricci scalar and metric determinant,
$\Phi$ is the gravitational scalar field, 
 $\Psi_m$ collectively describes the matter degrees of freedom, and where we have set $\hbar=c=1$. The dimensionless coupling constant $\omega$ regulates the deviations away from GR, to which FJBD theory reduces for $\omega\to\infty$. This can be seen more clearly  by performing the conformal transformation $\tilde g_{\mu\nu}=\Phi^{-1}\, g_{\mu\nu}$, where  the metric $g_{\mu\nu}$ is often referred to as Einstein-frame metric (as opposed to the Jordan-frame metric  $\tilde g_{\mu\nu}$). This transformation, together with the redefinition
 \begin{equation}\label{conf}
     \Phi=\exp\left(\sqrt{2}\,\alpha\, \frac{\varphi}{M_{\rm Pl}}\right)\,,\quad \alpha=\frac{1}{\sqrt{3+2\omega}} \,,
 \end{equation}
 allows for writing the Einstein-frame action~\cite{PhysRevD.1.3209}
 \begin{equation}
\label{einframe}\!
\!S=\!\!\int\!d^4 x \sqrt{-g} \left( \frac{M_{\rm Pl}^2}{2}R - \frac{1}{2}g^{\mu\nu} \partial_\mu\varphi \partial_\nu\varphi
\right) \! +S_m\!\!\left[\frac{g_{\mu\nu}}{\Phi(\varphi)},\Psi_m\right],
\end{equation}
with $R$ the Ricci scalar built from $g_{\mu\nu}$. From this action, it is clear that for $\omega\to\infty$ one simply obtains GR with a minimally coupled scalar field.
 
 The problem with FJBD theory is that it is strongly constrained by 
 solar system experiments, and in particular by the Cassini measurement of the Shapiro time delay, which bounds $\omega> 40000$ at 2$\sigma$ level~\cite{2003Natur.425..374B,Will:2014kxa}. This constraint renders the viable FJBD
theories very fine-tuned and close to GR, limiting their interest for cosmology. 
FJBD theory, however, is not the most general ST theory that one can conceive. Besides
adding a potential and making the coupling constant $\omega$ a function of $\Phi$ [or, equivalently,
considering a more general conformal factor than Eq.~\eqref{conf}], which
can already give rise to non-trivial phenomenology~\cite{Damour:1993hw,Barausse:2012da,Palenzuela:2013hsa,ST3,ST4,Khoury:2003rn,Hinterbichler:2010es}, one
can also generalize the action \eqref{einframe}
to include the Horndeski~\cite{Horndeski:1974wa}, beyond-Horndeski~\cite{Gleyzes:2014dya} and degenerate higher-order ST (DHOST) terms~\cite{Langlois:2015cwa, Crisostomi:2016czh}. This results in the cubic DHOST action derived in~\cite{BenAchour:2016fzp}, which describes the most general ST theory with no Ostrogradski ghosts.
When these additional terms are included, the phenomenology of ST theories becomes richer and more complex. 
In particular, several theories in the DHOST class possess a non-linear screening mechanism, whereby the local dynamics matches GR (thus evading the Cassini bound), while on large (cosmological) scales the scalar field
dynamics is left relatively unconstrained, thus possibly playing a role in the phenomenology of dark energy.

Several screening mechanisms have been proposed in the literature, ranging from chameleon/symmetron screening~\cite{Khoury:2003rn,Hinterbichler:2010es}, to the Vainshtein mechanism~\cite{Vainshtein:1972sx,Babichev:2013usa}, to kinetic screening (also known 
as $k$-mouflage)~\cite{Babichev:2009ee}. 
Among these, the latter is the only one evading constrains from
the speed of gravitational waves (GWs) measured by GW170817~\cite{Monitor:2017mdv,TheLIGOScientific:2017qsa},
the decay of GWs into the scalar mode \cite{Creminelli:2018xsv,Creminelli:2019nok},
instabilities of the scalar field induced by GWs \cite{Creminelli:2019kjy};  see also
e.g. \cite{Burrage:2017qrf} for bounds on chameleon/symmetron screening.
 Remarkably, the action giving rise to kinetic screening is also a very 
 simple generalization of the FJBD action \eqref{einframe}, which
 is  modified by making the kinetic term non-linear. In more detail, the 
 resulting action (often referred to as $k$-essence action) is given by~\cite{Chiba:1999ka, ArmendarizPicon:2000dh}
\be
S=\int \mathrm{d}^{4}x\sqrt{-g}\left[\frac{M_{\mathrm{Pl}}^{2}}{2}R + K(X)  \right]  + S_{m}\left[\frac{g_{\mu\nu}}{\Phi(\varphi)},\Psi_m\right] \,, \label{action}
\ee
where  $X\equiv g^{\mu\nu} \partial_\mu\varphi \partial_\nu\varphi$ and where we consider only the lowest order terms
\be
K(X)=-\frac{1}{2}X+\frac{\beta}{4\Lambda^4}X^2-\frac{\gamma}{8\Lambda^8}X^3+\ldots\;.\label{kessence}
\ee
Here, $\Lambda$ is the strong-coupling scale of the effective field theory, $\beta$ and $\gamma$ are dimensionless coefficients of $\mathcal{O}(1)$, and the conformal coupling
$\alpha$ [cf. Eq.~\eqref{conf}] can be $\sim\mathcal{O}(1)$, because the
kinetic screening allows for escaping the Cassini bound~\cite{terHaar:2020xxb}.

The validity of screening mechanisms, including kinetic screening, has only been studied in static and weak field regimes (e.g.~\cite{Khoury:2003rn,Hinterbichler:2010es,Babichev:2009ee,Babichev:2010jd,Babichev:2013pfa, Crisostomi:2017lbg}) or in quasistatic ones (e.g.~\cite{deRham:2012fg,deRham:2012fw,Dar:2018dra,Brax:2017wcj,deAguiar:2020urb}), and it has never been proven in the highly dynamical and non-linear regimes characterizing systems of compact objects, which can only be described by full-fledged numerical relativity simulations. 
In fact, even in the simple case of $k$-essence, for which the Cauchy problem is locally well-posed, pathologies arise in dynamical evolutions, with the field equations potentially changing character from hyperbolic to parabolic~\cite{Bernard:2019fjb,Bezares:2020wkn}. This change of character renders initial-value evolutions  unstable (i.e. ill-posed), but can be avoided 
in specific subclasses of $k$-essence theories, including ones giving kinetic screening~\cite{Bezares:2020wkn,terHaar:2020xxb}. Nevertheless, we showed in Ref.~\cite{terHaar:2020xxb} that when evolving neutron stars in these theories with kinetic screening, even though the equations always remain hyperbolic, the 
characteristic speeds of the scalar field may diverge when gravitational collapse is triggered. This happens also in vacuum close to critical collapse~\cite{Bezares:2020wkn}
and is at the very least a practical problem, as it makes the theory unpredictive [because
simulations cannot be evolved past this divergence as a result of the Courant–Friedrichs–Lewy (CFL) condition]. Moreover, it might also constitute a conceptual pathology, since the characteristic speeds
generalize the background scalar speed to non-linear orders, albeit in a gauge-dependent way~\cite{Bezares:2020wkn}.

In this paper, we build on the framework of Refs.~\cite{Bezares:2020wkn,terHaar:2020xxb} and show that this divergence of the characteristic speeds can be resolved by slightly modifying the dynamics by
adding a ``fixing equation'' in the spirit of the proposal by 
 Cayuso, Ortiz and Lehner~\cite{Cayuso:2017iqc} (see also \cite{Allwright:2018rut,Cayuso:2020lca}), which was in turn inspired by
 the work of Israel and Stewart on relativistic dissipative hydrodynamics~\cite{Israel:1979wp}.
 The addition of this equation modifies the dynamics of the theory, but 
 the true evolution of $k$-essence is recovered in the limit when a free timescale $\tau$,
 appearing in the fixing equation, vanishes. 
  In this work, we show that by
 taking a small $\tau\neq0$, the evolution of collapsing neutron stars (in spherical symmetry)
 matches the results of pure $k$-essence before the divergence of the characteristic speeds, but also proceeds unobstructed past it.
 We then use this framework to confirm the validity of kinetic screening in these dynamical settings and to study gravitational collapse in $k$-essence (in addition to non-linear stellar oscillations, for which a fixing equation is not needed).
 
  Remarkably, we find that 
 kinetic screening remains valid in oscillating stars (whose monopole scalar emission
 is suppressed to undetectable level for theories of interest for cosmology),
 while it  seems to break in collapsing systems. 
  In fact, our results suggest that 
  collapsing stars must shed away all their scalar hair before forming
   black holes, thus producing bursts of
   scalar radiation.
   These bursts are characterized by frequencies too low to be targeted by ground-based interferometers (at least for theories of interest for cosmology), but
   we conjecture that they may be detected by space-borne detectors such as LISA,
   if a supernova explodes in the Galaxy.

 In more detail, the paper is organized as follows. In 
 Sec.~\ref{fieldeqs} we present the field equations of $k$-essence. In 
 Sec. \ref{sectionID} we present  details on the static spherically symmetric  stars 
 of Ref.~\cite{terHaar:2020xxb}, and use them to review the kinetic screening mechanism.
 The numerical setup for our simulations, including the fixing equation, is described in Sec.~\ref{dynamicsection},
 where we also present results for the dynamical evolution of
 oscillating and collapsing neutron stars. 
  In Sec. \ref{conclusions} we draw our conclusions. 
Throughout this paper we assume a metric signature $(-+++)$ and units where $\hbar=c=1$. In the Appendix we review the relation between
these units and the units $G=c=M_{\odot}=1$ that need to be used to
simulate the dynamics of neutron stars, and explain why studying numerically stars in
$k$-essence theories of relevance for cosmology is challenging as a result
of the hierarchy of scales involved.

\section{The field equations of $k$-essence theories}\label{fieldeqs}
By varying the $k$-essence action \eqref{action}, one obtains
the equations of motion for the metric and scalar field
\begin{align}\label{fieldeqs1}
&G_{\mu\nu}=8 \pi G\big( T^{\varphi}_{\mu\nu} + T_{\mu\nu}\big)\;,\\\label{fieldeqs2}
%%%%%%%%%%%%%%%%%%%%%
&\nabla_\mu\left[K'(X) \nabla^\mu \varphi\right]=\frac{1}{2}\mathcal{A} T\;,
\end{align}
where $G_{\mu\nu}$ is the Einstein tensor constructed from
the Einstein-frame metric $g_{\mu\nu}$, we define $\mathcal{A} \equiv -\Phi'(\varphi)/[2\Phi(\varphi)]$, and the scalar field and matter energy-momentum tensors are defined as
\begin{eqnarray}
T^{\varphi}_{\mu\nu} &=& K(X)g_{\mu\nu}-2K'(X) \de_\mu \varphi \de_\nu \varphi\; ,\\
%%%%%%%%%%%%%%%%%%%%%
T_{\mu\nu} &=& \frac{2}{\sqrt{-g}}\frac{\delta S_{m}}{\delta g_{\mu\nu} }\;,    \label{SET}
\end{eqnarray}
 with
%$\varphi_\mu=\nabla_\mu \varphi$ and
$T=T_{\mu\nu}g^{\mu\nu}$. Although the Einstein frame is convenient when solving these equations numerically, we convert back to the Jordan frame (e.g. the frame in which matter follows geodesics) to present and interpret our results. 
 
To solve this system of coupled equations, we make a few assumptions. 
First, we model matter by a perfect fluid in the Jordan frame, with
  rest-mass density $\tilde{\rho}_0$, specific internal energy $\tilde{\epsilon}$, pressure $\tilde{P}$ and four-velocity $\tilde{u}_{\mu}$.
From the definition \eqref{SET},
the stress energy tensor in the Einstein frame, $T^{\mu\nu}$,
is related to the one in the Jordan frame, $\tilde{T}_{\mu\nu}$, by
$T^{\mu\nu}=\tilde{T}^{\mu\nu}\Phi^{-3}$, $T_{\mu\nu}=\tilde{T}_{\mu\nu}\Phi^{-1}$~\cite{Barausse:2012da,Palenzuela:2013hsa}.
We can then write the Einstein frame stress energy tensor as
\begin{equation}
T_{\mu\nu} = [\rho_0(1+\epsilon) + P]u_{\mu}u_{\nu} + P\,g_{\mu\nu}\,,
\end{equation}
with $\rho_0$,  $\epsilon$, $P$ and $u_{\mu}$
related to their Jordan frame counterparts by~\cite{Barausse:2012da,Palenzuela:2013hsa}
$u^\mu=\tilde u^\mu\, \Phi^{-1/2}$ (which ensures that the four-velocity has unit norm in both frames), $P=\tilde P \, \Phi^{-2}$ and $\rho_0=\tilde\rho_0\,\Phi^{-2}$. 
These relations imply that if one considers an equation of state relating
 $\tilde{\rho}_0$,  $\tilde{\epsilon}$, $\tilde{P}$ in the Jordan frame, the corresponding equation of state in the Einstein frame will also (in general) involve the scalar field via the conformal factor~\cite{Barausse:2012da,Palenzuela:2013hsa}.

Since in the Jordan frame  matter is not directly coupled to the
scalar field but only to the metric, the usual conservation laws 
of the matter stress energy tensor and baryon number apply in that frame.
Transforming those conservation laws to the 
 Einstein frame one obtains 
 \begin{eqnarray}
\nabla_{\mu}T^{\mu\nu} &=&\mathcal{A}~\nabla^\nu \varphi \,T \,,
\label{fieldeqs4}
\\
\nabla_{\mu}(\rho_0 u^{\mu}) &=& \rho_0\mathcal{A}u^\mu \nabla_\mu\varphi   \,. \label{fieldeqs3}
\end{eqnarray}
Therefore, unlike in the Jordan frame, the stress energy tensor and the baryon number are {\it not} conserved in the Einstein frame.

\section{Static Solutions}\label{sectionID}

In this section, we assume a spherically symmetric and static ansatz
for both the metric and the fluid, with the goal of finding solutions
representing isolated stars and probing the validity of the kinetic screening mechanism.
We use areal coordinates for the metric, for which we adopt the ansatz
\be
\mathrm{d} s^2=g_{tt}(r)\mathrm{d} t^2+g_{rr}(r)\mathrm{d}r^2+r^2\mathrm{d}\Omega^2\;,
\ee
with $\mathrm{d}\Omega$ the solid angle element. (Note that
this gauge differs from the one we will adopt in  Sec.~\ref{dynamicsection}
to study the time evolution of these objects, although transforming between the two is straightforward.)

Although screening solutions exist in $k$-essence for any $\beta<0$, $\gamma>0$ in equation (\ref{kessence})\footnote{Unfortunately, for this choice of the parameter signs, $k$-essence does not admit a standard (Wilsonian) UV completion. For this reason an alternative approach, such as the ``fixing equation'' method that we utilize in this paper, is necessary when the characteristic scalar speeds diverge.}, in the following we set $\beta=0$ and $\gamma=1$. This ensures
that the theory satisfies
the condition  $1 +  2\, X\, K''(X)/K'(X) >0$ for all $X$ \cite{Bezares:2020wkn, Babichev:2007dw, Brax:2014gra}, which in turn implies that
the field equations remain always strongly hyperbolic (thus allowing
us to study the Cauchy problem in Sec.~\ref{dynamicsection}). The results presented in this work, however, hold (qualitatively) for more general $\beta$ and $\gamma$, provided that the above condition is satisfied.

We wrote a Tolman–Oppenheimer–Volkoff (TOV) solver in Mathematica~\cite{Mathematica} to find these static spherical stars, imposing regularity at the origin by solving the field equations perturbatively at small radii. 
These perturbative results are then used as initial data for an outbound integration (in the radial coordinate) starting at small but nonzero $r$. The initial data also depends on 
the central density $\rho_c$ and on the
central value of the scalar field, which is fixed through a shooting procedure by requiring $\varphi\rightarrow\varphi_\infty$ (with $|\varphi_\infty|/\Lambda\lesssim10^{-3}$) as $r\rightarrow\infty$.

To close the system, we consider a polytropic equation of state $\tilde{P}=K\tilde{\rho}_0^\Gamma$, $\tilde{P}=(\Gamma-1)\tilde{\rho}_{0}\tilde{\epsilon}$ in the Jordan frame. We will mainly be studying neutron stars, and use $K=123\;G^3M^2_\odot/c^6$ and $\Gamma=2$. Instead, when studying weakly-gravitating stars such as the Sun, we consider $K=5.9\times10^{-5}\;G^{1/3}R^{2/3}_\odot/c^{2/3}$ and $\Gamma=4/3$.

\subsection{Screening in isolated stars}
\label{screening}
As  clear from the $k$-essence action (\ref{kessence}), non-linearities in $X$ are suppressed by the physical scale $\Lambda$. 
If we assume that the scalar field is responsible for dark energy (DE), $\Lambda$ needs to be of the order of $\Lambda_\mathrm{DE}\sim(H_0 M_\mathrm{Pl})^{1/2}\sim 2 \times 10^{-3}\;\mathrm{eV}$, where $H_0$ is the present-day Hubble expansion rate. 
At spatial infinity, the theory is in the perturbative regime
and behaves as FJBD theory. However, at a ``screening radius''
 $r_k\sim \Lambda^{-1}\sqrt{M/M_\mathrm{Pl}}$, with $M$ the mass of the star,
 the non-linear terms start dominating, suppressing (or ``screening'') scalar effects at $r\lesssim r_k$. Within this screening radius, $k$-essence is equivalent to GR. 
In this section, we show explicitly how the screening mechanism in $k$-essence affects the gravitational force. We evaluate the latter as a function of the Jordan frame radius, and we will be especially interested in the regimes $\tilde{r}_\star<\tilde{r}<\tilde{r}_k$ (where screening is at work; $\tilde{r}_\star$ being the radius of the star), and $\tilde{r}>\tilde{r}_k$ (where $k$-essence starts deviating from GR).

\begin{figure*}
    \centering
    \includegraphics[height=0.32\textwidth]{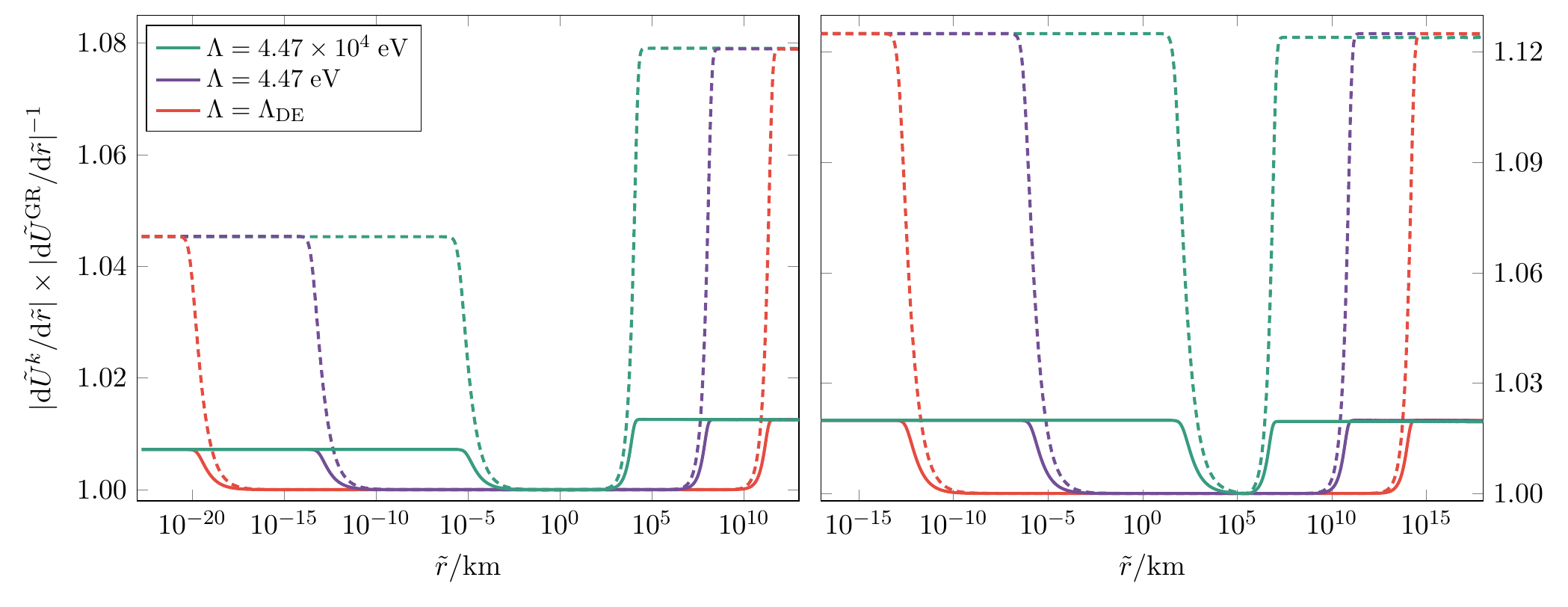}
    \caption{Deviations of the Newtonian acceleration from GR for neutron stars (left) and Sun-like stars (right), and for different values of $\Lambda$ in $k$-essence. We consider both $\alpha\approx 0.14$ (solid lines) and $\alpha\approx 0.35$ (dashed lines).}
    \label{fifthforceplot}
\end{figure*}

The screening mechanism aims to suppress  the scalar fifth force on local scales, and thus tends to make the gravitational force inside the screening radius equal to the one in GR.
Since the Newtonian potential $\tilde{U}$ is encoded in the fall-off of the Jordan-frame metric
component $\tilde{g}_{tt}$ far from the star, $\tilde{U}\approx -(\tilde{g}_{tt}+1)/2$,
we can quantify the difference between the ``Newtonian acceleration'' $|{\rm d} \tilde{U}/{\rm d} \tilde{r}|$
in GR and $k$-essence. In Fig.~\ref{fifthforceplot}, we show the  ratio of these two accelerations for six different solutions: three neutron stars (left panel) and three Sun-like stars (right panel). To generate these solutions, we have considered three different values for the strong-coupling scale $\Lambda=\{4.47\times10^4\;\mathrm{eV},\;4.47\;\mathrm{eV},\;\Lambda_\mathrm{DE}\}$, and considered two different values for the conformal coupling constant $\alpha$. For neutron stars, the central density is fixed to $\rho_c=9.3\times10^{14}\;\gcc$, whereas for Sun-like stars the central density is fixed to $\rho_c=77\;\gcc$. With fixed $\rho_c$, $\alpha$, and $\varphi_\infty$, we expect the central value of the scalar field (which has dimensions of an energy) to go as 
\begin{eqnarray}
\varphi_c\propto \Lambda\;,
\label{relphi}
\end{eqnarray}
a relation that is indeed satisfied by our static solutions (at least for sufficiently small  $\Lambda$ giving rise to kinetic screening), as we have explicitly verified.
We stress that producing stellar solutions
with $\Lambda\approx\Lambda_\mathrm{DE}$
is far from trivial. In order to resolve the interior of the star, which
is crucial to impose regularity at the center (cf. also \cite{terHaar:2020xxb})
one needs to use internal code units adapted to the problem (e.g. $G=c=M_\odot=1$ or $G=c=R_\odot=1$). Converting $\Lambda_{\rm DE}$ to these units yields very small values $\Lambda_{\rm DE}\sim 10^{-12}$ (see the Appendix \ref{app}), which are difficult to handle.
We also stress that this is an issue  due to the hierarchy of scales in the problem (which
involves both local stellar scales and the cosmological scale $\Lambda_{\rm DE}$), and which is therefore independent of the choice of units.

As can be seen from Fig.~\ref{fifthforceplot}, the screening works in a similar way in
 Sun-like and neutron stars.  At radii larger than $\tilde{r}_k$, the $k$-essence Newtonian acceleration deviates from the one in GR, with the magnitude of the deviation depending on the value of $\alpha$ (in Fig.~\ref{fifthforceplot} the solid lines correspond to $\alpha\approx0.14$ and the dashed lines to $\alpha\approx0.35$). However, when the radius reaches $\tilde{r}_k$, the fifth force starts being suppressed, and $|\mathrm{d}\tilde{U}^k/\mathrm{d}\tilde{r}|\times |\mathrm{d}\tilde{U}^\mathrm{GR}/\mathrm{d}\tilde{r}|^{-1}$ gets very close to unity. As expected, the smaller the strong-coupling scale $\Lambda$, the larger the screening radius $\tilde{r}_k$ within which the fifth force is suppressed. Finally, deep inside the star the fifth force reappears as well. This is expected because at the center of the star the kinetic energy of the scalar field $\tilde{X}$ vanishes because of regularity, and thus $k$-essence reduces to FJBD theory (cf. also \cite{terHaar:2020xxb}).

To check whether these results hold also beyond Newtonian order, and more specifically at the
first post-Newtonian order (1PN) that is tested in the solar system, we compare the exterior
of our numerical solutions to the
parametrized post-Newtonian (PPN) expansion~\cite{Will:1993hxu,Will:2014kxa}, and extract the PPN parameters $\beta^\mathrm{PPN}$ and $\gamma^\mathrm{PPN}$ (which are unity in GR). The latter are defined in our areal coordinates as~\cite{Barausse:2014tra}
\begin{align}\label{pn1}
    &\tilde{g}_{tt}(\tilde{r})=-1+\frac{2G\tilde{M}}{\tilde{r}}\nonumber\\&\qquad-2\left(\beta^\mathrm{PPN}-\gamma^\mathrm{PPN}\right)\left(\frac{G\tilde{M}}{\tilde{r}}\right)^2+\mathcal{O}\left(\tilde{r}^{-3}\right)\;,\\
   & \tilde{g}_{\tilde{r}\tilde{r}}(\tilde{r})=1+2\gamma^\mathrm{PPN}\frac{G \tilde{M}}{\tilde{r}}+\mathcal{O}\left(\tilde{r}^{-2}\right)\;.\label{pn2}
\end{align}
For this analysis, we consider only $k$-essence theories of cosmological relevance, and thus take $\Lambda=\Lambda_\mathrm{DE}$ (while fixing $\alpha\approx0.14$).
We extract the PPN parameters from solutions for a Sun-like star in the regime where $\tilde{r}_\star < \tilde{r} < \tilde{r}_k$ and compare their values to the constraints from solar system tests. This is justified because solar system experiments are performed well within the screening radius of the Sun,
but it also poses a practical problem. Inside the screening radius, the non-linear terms in the action are important, and one cannot simply perform a naive perturbative PN expansion of the metric and scalar field~\cite{starsNohair1}. 
This is evident from the fact that only outside the screening radius does the
scalar field decay as $1/\tilde{r}$ (in orders of which the PN expansion would be performed). Equivalently, one can observe that a naive PN expansion
would lead to the wrong conclusion that at leading  (i.e. 
Newtonian) order $k$-essence should reduce to FJBD (which is clearly not the case inside $\tilde{r}_k$). We therefore
use our numerical solutions and simply fit them with the ansatz \eqref{pn1}--\eqref{pn2} to extract $\gamma^\mathrm{PPN}$ and $\beta^\mathrm{PPN}$, obtaining
 \begin{align}
    \gamma^\mathrm{PPN}-1&=(-5.54\pm1.68)\times 10^{-10}\;,\\
    \beta^\mathrm{PPN}-1&=(1.27\pm0.733)\times10^{-3}\;,
 \end{align}
 where the error bars are at $1\sigma$.
The PPN parameters are constrained close to unity by solar system observations~\cite{2003Natur.425..374B,Will:2014kxa}, with bounds $|\gamma^\mathrm{PPN}-1|,\,|\beta^\mathrm{PPN}-1|\lesssim 10^{-5}$.
As can be seen, our results are therefore compatible with these bounds
at $2\sigma$ level, but our statistical error on $\beta^\mathrm{PPN}-1$
is much larger than the experimental bounds. This is because 
it is challenging to extract $\beta^\mathrm{PPN}$ from
our numerical solutions, since it appears at higher order than $\gamma^\mathrm{PPN}$
in Eqs.~\eqref{pn1}--\eqref{pn2}. This problem is also exacerbated by the low compactness 
of the Sun, which limits the range of radii on which we can perform our fit.
Repeating indeed the procedure for more compact stars (e.g. for neutron stars), we find the more
precise result
 \begin{align}
    \gamma^\mathrm{PPN}-1&=(-2.98\pm 1.38)\times 10^{-12}\;,\\
    \beta^\mathrm{PPN}-1&=(1.10\pm 0.764)\times10^{-10}\;,
 \end{align}
 which is again in perfect agreement with the experimental bounds.

\subsection{Mass-Radius Curves}
\label{MR}

To study the screening mechanism in a dynamical setting, in Sec.~\ref{dynamicsection} we will evolve screened neutron stars in $k$-essence. In order to have a better understanding of the characteristics of these stars, we first take a closer look at their mass $\tilde{M}$ and radius $\tilde{r}_\star$. 
When screening is at play, however, the definition of mass is subtle.
Although the gravitational mass is formally defined at spatial infinity, in practice the masses of stars are measured by the observation of orbital motion of bodies/gas well inside the screening radius. Therefore, we can define two different masses, one at spatial infinity ($\tilde{M}_\infty$) and one ``felt'' by bodies surrounding the star but located well inside its screening radius ($\tilde{M}_{\rm screened}$). 
In practice, one can extract the former from the metric component $\tilde{g}_{tt}\approx -1+2G\tilde{M}/\tilde{r}$ at spatial infinity, and the latter by fitting it in the range $\tilde{r}/(GM_\odot)\sim10^5$--$10^7$, which is the typical separation e.g. of binary pulsar systems.

\begin{figure}[h]
    \centering
    \includegraphics[height=0.32\textwidth]{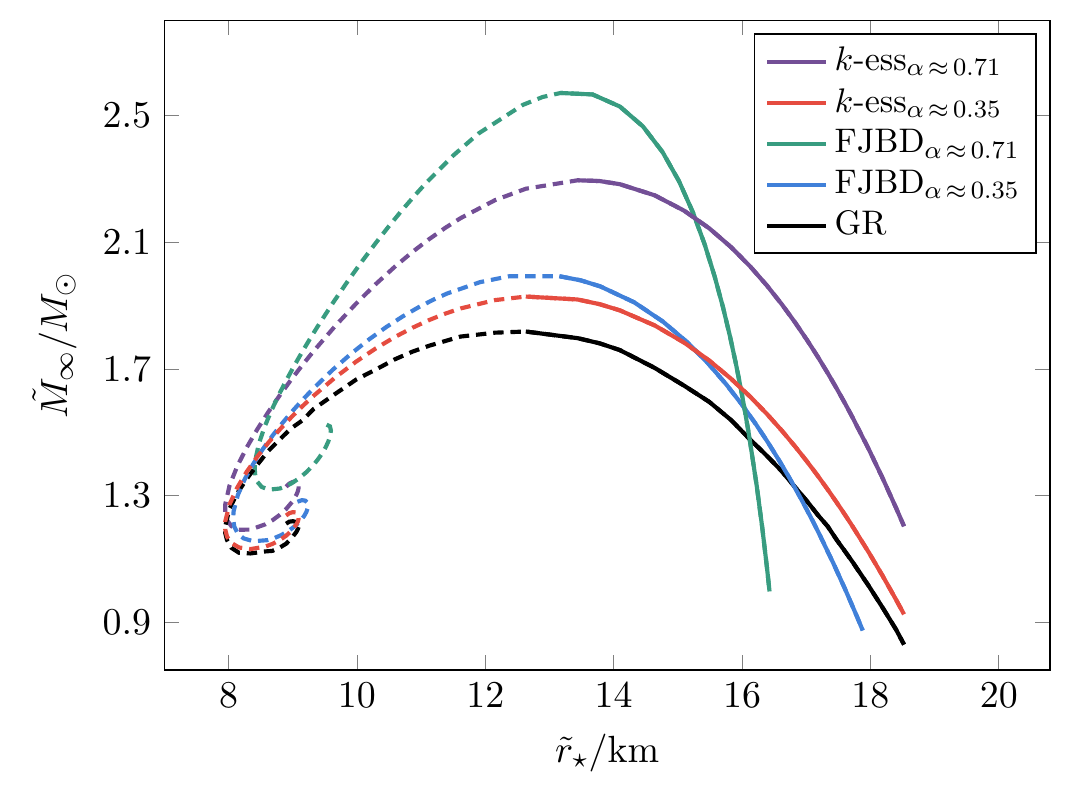} 
    \caption{Mass-radius curves for $\tilde{M}=\tilde{M}_\infty$ in $k$-essence (with $\Lambda=\Lambda_\mathrm{DE}$), FJBD theory, and GR. We have fixed $\alpha\approx0.35$ and $\alpha\approx0.71$, and vary the central density to generate different stars (following the curves from right to left corresponds to increasing $\rho_c$). We have differentiated between stable (solid lines) and unstable branches (dashed lines).}
    \label{mrcurvesplot}
\end{figure}

Let us start by considering the mass at spatial infinity.
First, we fix the central density to $\rho_c=9.3\times10^{14}\,\gcc$, and consider three different values for the conformal coupling constant $\alpha$. The corresponding neutron star masses and radii in GR, $k$-essence, and FJBD are listed in Table~\ref{mrtable}. Then, we consider a range of central densities to generate different stars in the same three theories, while fixing $\alpha\approx0.35$ and $\alpha\approx0.71$, and show the mass-radius curves in Fig.~\ref{mrcurvesplot}.

\begin{table}
\begin{center}
 \begin{tabular}{c | c | c | c} 
 \hline \hline
& $\alpha$ & $\tilde{M}_\infty/M_\odot$ & $\tilde{r}_\star/\rm km$\\
 \hline \hline
 GR & absent & 1.719 & 14.47 \\[0.5ex]\hline
 \multirow{3}{*}{$k$-essence} & $0.14$ & 1.741 & 14.47 \\
 & $0.35$ & 1.855 & 14.47 \\
  &$0.71$ &2.262 & 14.47 \\[0.5ex]\hline
 \multirow{3}{*}{FJBD} & $0.14$ & 1.752 & 14.42 \\
 & $0.35$ & 1.929 & 14.16 \\
  & $0.71$ & 2.572 & 13.51 \\[0.5ex]
  \hline\hline
\end{tabular}
\captionsetup{width=.8\linewidth}
\caption{Neutron star solutions for a central density of $\rho_c=9.3\times10^{14}\;\gcc$ in GR, $k$-essence ($\Lambda=\Lambda_\mathrm{DE}$), and FJBD theory.}
\label{mrtable}
\end{center}
\end{table}

In Table~\ref{mrtable}, we show that  both $k$-mouflage and FJBD stars become heavier when $\alpha$ increases. 
Their masses also deviate from the masses of the GR solutions, as expected. Indeed, the gravitational mass
is extracted at spatial infinity, where
no screening is present and scalar effects can be significant. Conversely, the radius of the star $\tilde{r}_\star$ (defined
by $\tilde{P}(\tilde{r}_\star)=0$) is within the screening radius, 
and we therefore find that in $k$-essence it matches the GR stellar radius. As the fifth force is not screened in FJBD theory, 
stellar radii in the latter do show differences from $k$-essence and GR. In Fig.~\ref{mrcurvesplot}, we show the mass-radius curves for the three theories, and find that deviations from the GR mass-radius curve are more pronounced for larger $\alpha$ in both $k$-essence and FJBD theory. 

\begin{figure}[h]
    \centering
    \includegraphics[height=0.32\textwidth]{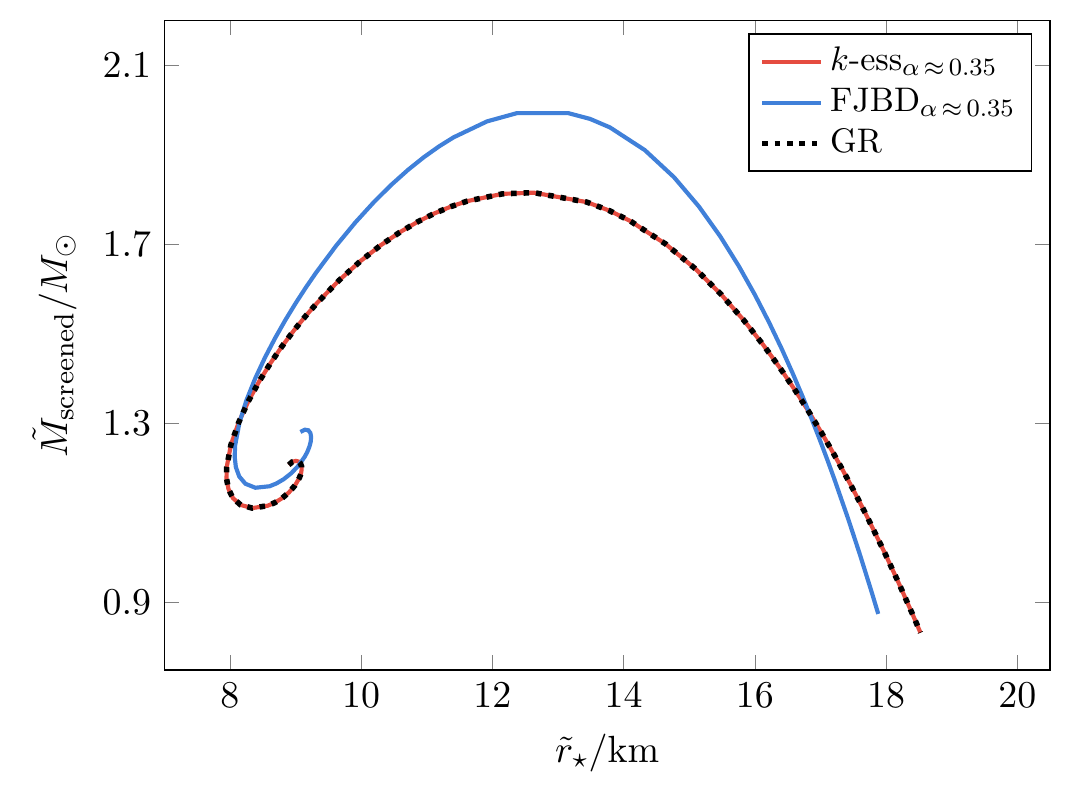}   
    \caption{Mass-radius curves for $\tilde{M}=\tilde{M}_{\rm screened}$ in $k$-essence (with $\Lambda=\Lambda_\mathrm{DE}$), FJBD theory, and GR. We have fixed $\alpha\approx0.35$, and vary the central density to generate different stars.}
    \label{mrcurvesfiniteplot}
\end{figure}

Let us now consider the screened mass $\tilde{M}_{\rm screened}$.
The resulting mass-radius curves can be found in Fig.~\ref{mrcurvesfiniteplot}. One can see that there is a perfect overlap between the GR and $k$-essence curves. This makes sense since we are fitting the mass within the screening radius $\tilde{r}_k$, where the two theories are equivalent. We do instead find deviations for the FJDB curve, since there is no screening in that theory.

\subsection{Scalar charges and scalar field energy}
In gravitational theories that modify/extend GR, the universality of free fall (which in GR
is satisfied as the theory obeys the equivalence principle) is typically violated, at least for strongly gravitating objects such as neutron stars~\cite{Eardley,Will:1989sk,Damour:1992we,Damour:1993hw,Barausse:2012da,Palenzuela:2013hsa,ST3,ST4,Yagi:2013ava,Yagi:2013qpa,Mirshekari:2013vb,Gupta:2021vdj} and black holes~\cite{Yagi:2015oca,Barausse:2016eii,starsNohair1}. This amounts to a violation of the strong equivalence principle and is ripe of consequences for gravitational-wave generation, as it gives rise to dipole gravitational emission from binary systems (and even monopole emission, for non-circular binaries and collapsing stars), as well
as to modifications in the conservative dynamics of binaries~\cite{Eardley,Will:1989sk,Damour:1992we,Mirshekari:2013vb,Yagi:2013ava}.

Violations of the strong equivalence principle in modified gravitational theories are usually parametrized by ``sensitivities'' or ``charges'', i.e. additional ``hair parameters''
describing compact objects and their effective coupling to the non-tensor gravitons 
that are generally present in these theories. These charges vanish in the low-compactness limit
if the matter fields couple minimally to the metric [as is the case for the ST theories that we consider, cf. the Jordan-frame action \eqref{Jframe_action}], i.e. if the weak equivalence principle is satisfied. However, they can be significant for neutron stars or black holes, especially if non-linear phenomena (e.g. ``scalarization'') are at play~\cite{Damour:1993hw,Barausse:2012da,Palenzuela:2013hsa,ST3,ST4,Silva:2017uqg,Silva:2018qhn,Dima:2020yac}. 

In ST theories, one can indeed define a dimensionless scalar charge $\bar{\alpha}$
describing the effective coupling between the scalar field 
and compact objects.
From the decay of the scalar field near spatial infinity, 
\begin{equation}\label{falloff}
    \varphi=\varphi_\infty+\frac{\varphi_1}{r}+\mathcal{O}\left(\frac{1}{r^2}\right)\;,
\end{equation}
we can extract the scalar charge as~\cite{Damour:1992we,Palenzuela:2013hsa}
\begin{equation}\label{chargedef}
    \bar{\alpha}=\sqrt{\frac{4\pi} {G}}\frac{\varphi_1}{M_\infty}\;,
\end{equation}
with $M_\infty$ the gravitational mass in the Einstein frame, extracted from the asymptotic expansion $g_{tt}=-1+2G M_\infty/r+...\;$ at spatial infinity.
As mentioned above, the importance of these scalar charges lies
in the modifications that they induce on gravitational-wave generation. Non-zero charges
can produce monopole and dipole radiation (the former only in eccentric binaries), as
opposed to the quadrupole emission of GR (which also gets modified by the scalar charges)~\cite{Damour:1992we,Will:1989sk,Mirshekari:2013vb}.
Scalar charges may also modify the conservative dynamics of binary systems with respect to GR~\cite{Damour:1992we,Will:1989sk,Mirshekari:2013vb}. 
As a result, non-zero scalar charges can provide a way to test the theory experimentally, a program that was indeed pursued in FJBD-like theories~\cite{Freire:2012mg}.

\begin{figure}[h]
    \centering
    \includegraphics[height=0.32\textwidth]{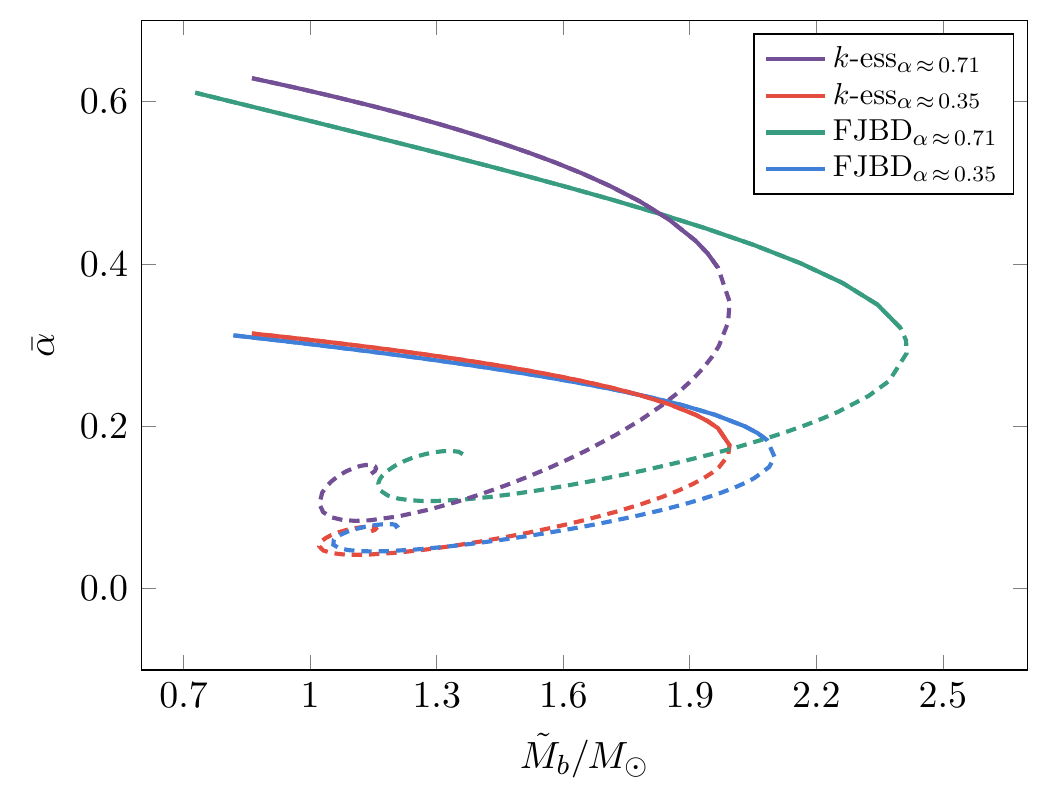}   
    \caption{The scalar charge $\bar{\alpha}$ as a function of the baryon mass $\tilde{M}_b$ for $k$-essence (with $\Lambda=\Lambda_\mathrm{DE}$) and FJBD theory for conformal coupling constants $\alpha\approx0.71$ and $\alpha\approx0.35$. Again, the stable branches are presented by solid lines, and the unstable branches by dashed lines.}
    \label{scalarchargeplot}
\end{figure}

Results for the scalar charges in $k$-essence and FJBD theory for two values of
the conformal coupling ($\alpha\approx0.71$ and $\alpha\approx0.35$) are shown in Fig. \ref{scalarchargeplot},
as functions of the bayon mass in the Jordan frame,
\begin{equation}
    \tilde{M}_b=\int \mathrm{d}^3\tilde{x} \sqrt{-\tilde{g}}\; \tilde{\rho}_0 \tilde{u}^0\;.
\end{equation}
We find that the scalar charge is of the same order of magnitude in $k$-essence and FJBD, with a larger $\alpha$ corresponding to larger $\bar{\alpha}$ in both theories (for a fixed central density). Another similarity between the theories is that by increasing $\rho_c$, the scalar charge decreases (i.e., as expected, the scalar charges 
decreases with compactness). 

Differences can be found in both the baryon mass and scalar charge  shown in Fig. \ref{scalarchargeplot}. 
While the baryon mass was expected to behave differently
in $k$-essence and FJBD theory (since it is defined 
inside the screening radius), the behavior of the scalar charge is at first sight surprising.
Just like the gravitational mass $M_\infty$, the scalar charge $\bar{\alpha}$ is a quantity that is extracted near spatial infinity. In this regime there is no screening, and the linear terms of the scalar action (e.g. the FJBD terms) will dominate over the non-linear ($k$-essence) ones. Therefore, in the scalar sector, $k$-essence is equivalent to FJBD theory near spatial infinity, and 
one would expect the scalar charges to be the same in the two theories.
In fact, for fixed central density, the coefficient $\varphi_1$
that regulates the decay of the scalar field and which enters the definition \eqref{chargedef}
is the same in the two theories, but the Einstein frame mass [which also enters 
Eq.~\eqref{chargedef}] is not. 
As a result, the scalar charges are different.

An important caveat is that the scalar charge, being 
extracted from
the fall-off of the scalar field near spatial infinity,
describes the solution in a region
where no screening is present and $k$-essence behaves perturbatively.
It should be stressed, however, that the formalism to compute the impact of the scalar charges on gravitational-wave emission and on the conservative dynamics {\it also} uses PN theory, which is only valid
outside the screening radius. As pointed out by \cite{starsNohair1}, this limits the 
physical meaningfulness of the scalar charges, which are only relevant for the conservative and dissipative dynamics of binary systems with separations larger
than the sum of their screening radii. Since for   $\Lambda\approx \Lambda_{\rm DE}$
 a neutron star's screening radius is $\sim 10^{11}$ km, this excludes known
 binary pulsars,
  whose separation is typically  $\lesssim 10^6$ km. 
  
  Therefore, testing $k$-essence with 
  binary pulsar timing data would require solving for the non-linear dynamics inside the
  screening radius, and cannot rely on PN theory. While some work in this direction has been done by using a Wentzel-Kramers-Brillouin approximation~\cite{deRham:2012fw,deRham:2012fg,Dar:2018dra}, results are still inconclusive
  because full-fledged non-linear simulations of the dynamics of $k$-essence within the screening radius are still missing. We will contribute to solving this problem in a forthcoming publication. For the moment, let us stress two points.
  
  First, it should be noted that the  Square Kilometre Array (SKA) is expected to 
  discover several new millisecond pulsars, especially near the Galactic center~\cite{Janssen:2014dka}. Based on the distribution of semi-major axes of known S-stars (which are $\gtrsim 1000$ au $\approx 10^{11}$ km~\cite{Gillessen:2008qv}), it is not to be excluded that the conservative dynamics of millisecond pulsars
  around SgrA$^\star$ may be used, in the near future, to test $k$-essence in the perturbative regime where scalar charges are relevant. 
  Second, even though a complete formalism to describe the scalar charges and the dynamics of a binary at separations
  smaller than the screening radii of its components is currently missing, we
  expect scalar effects to be suppressed inside the screening radius. This is evident
  from the discussion of Secs. \ref{screening} and \ref{MR}, but we can also see it explicitly by calculating
  the contribution of the scalar field to the energy of the star.
  
The energy of the scalar field  can be defined as the spatial integral of the time component of the current $\tilde{J}^\mu=\tilde{T}^{\mu\nu}_\varphi n_\nu$, where $n^\mu=\delta^\mu_t/\sqrt{-\tilde{g}_{tt}}$ is the unit norm
vector orthogonal to the foliation. The scalar field energy (in the Jordan frame) within a radius $\tilde{r}$ is then 
\begin{align}
    \tilde{E}_\varphi(\tilde{r})&=-\int_{|\boldsymbol{x}|<\tilde{r}} \mathrm{d}^3 \tilde x \sqrt{-\tilde{g}}\;\tilde{J}^t \nonumber \\
 %   &=4\pi \int \mathrm{d}r \sqrt{g_{rr}} r^2 \alpha^{-2}T^\phi_{tt}\;,\nonumber \\
    &=4\pi \int_0^{\tilde{r}} \mathrm{d}\tilde{r}\;[-\tilde{r}^2\sqrt{\tilde{g}_{\tilde{r}\tilde{r}}} \Phi^2K(X)]\;,
\end{align}
where the minus sign ensures that $\tilde{E}_\varphi>0$.

\begin{figure*}
    \centering
    \includegraphics[width=\textwidth]{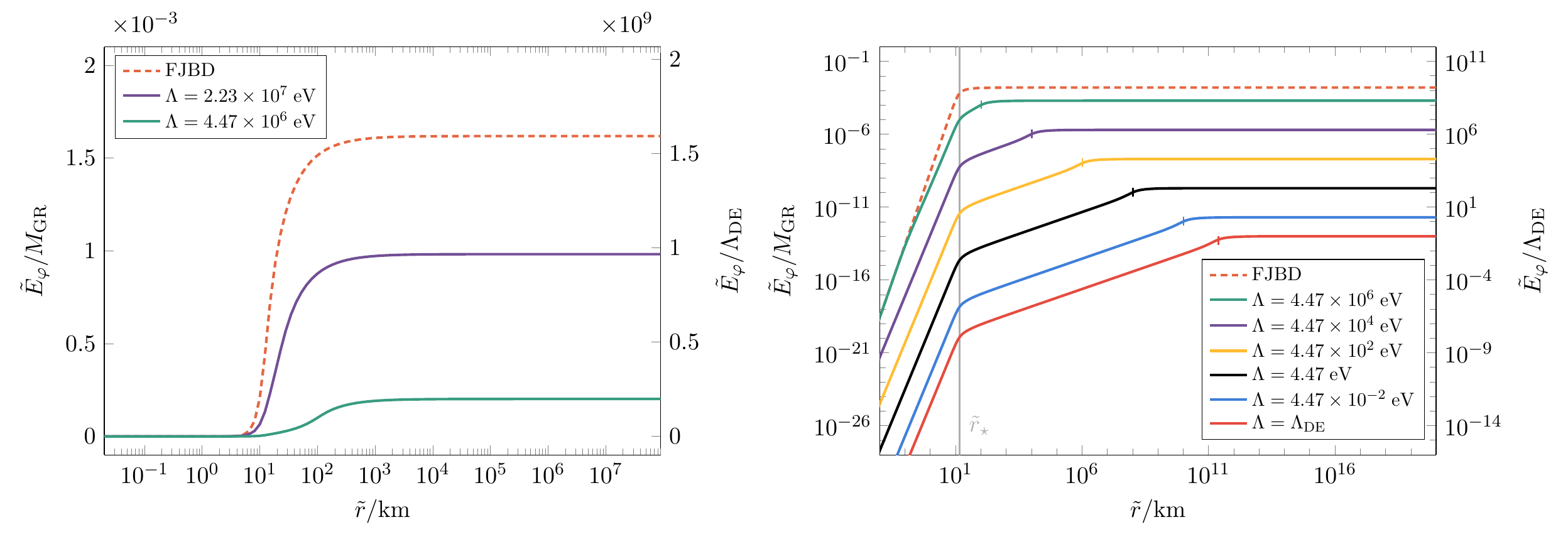}
    \caption{Left: The scalar energy as a function of the Jordan frame radius for a neutron star in FJBD theory, and in two $k$-essence theories (with two distinct strong coupling scales $\Lambda$). Right: The scalar energy of the solutions presented in Table~\ref{scalarenergytable}. The radius of the star $\tilde{r}_\star$ is indicated by a light gray line, and the screening radii $\tilde{r}_k$ by small vertical lines on top of the solutions.}
    \label{scalarenergyplot}
\end{figure*}

In Table~\ref{scalarenergytable}, we present seven different solutions for varying $\Lambda$
and report their total scalar field energy $\tilde{E}^\varphi_\infty$. Besides the value of $\tilde{E}^\varphi_\infty$, normalized by both $\Lambda_\mathrm{DE}$ and $M_\mathrm{GR}$ (equal to $1.719\;M_\odot$, see Table~\ref{mrtable}), we  also show the gravitational mass $\tilde{M}_\infty$, the baryon mass $\tilde{M}_b$, the radius of the star $\tilde{r}_\star$, and the screening radius $\tilde{r}_k$ of the solutions. All these quantities are evaluated in the Jordan frame. Note that $\Lambda=\infty$ corresponds to FJBD theory. In Fig.~\ref{scalarenergyplot}, we plot $\tilde{E}_\varphi(\tilde{r})$ for the solutions presented in Table~\ref{scalarenergytable}. 

There are a few things to notice in Table~\ref{scalarenergytable}. First, as expected, we see that the gravitational mass of the stars decreases with decreasing $\Lambda$ (and thus more suppression of the scalar field in the screened regime). At the same time, the radius of the stars increases, resulting in less compact stars for smaller $\Lambda$. We also confirm again that the screening radius increases for decreasing $\Lambda$. The scalar field energy at infinity is always small compared to the gravitational mass in GR (i.e. $\tilde{E}^\varphi_\infty/M_\mathrm{GR}\lesssim10^{-3}$), and for $\Lambda\sim10^{-1}\;\mathrm{eV}$ it starts being $\tilde{E}^\varphi_\infty/\Lambda_\mathrm{DE}\lesssim\mathcal{O}(1)$. In Fig.~\ref{scalarenergyplot}, we can see the scalar energy as a function of $\tilde{r}$. It starts being suppressed when screening kicks in (deep within the star, not included in the figure), and even more so once we go outside the surface of the star (indicated with a light gray line in the figure). When $\tilde{r}\sim \tilde{r}_k$, the profile flattens and the scalar field energy asymptotes to its value at infinity.

\begin{table*}
\begin{center}
 \begin{tabular}{c | c | c | c | c | c | c } 
 \hline \hline
$\Lambda$ & $\tilde{M}_\infty/M_\odot$ & $\tilde{M}_b/M_\odot$ & $\tilde{r}_\star/\rm km$ & $\tilde{r}_k/\rm km$ & $\tilde{E}^\varphi_\infty/\Lambda_\mathrm{DE}$ & $\tilde{E}^\varphi_\infty/M_\mathrm{GR}$ \\ 
 \hline\hline
$\infty$ & 1.752 & 1.889 & 14.42  & absent & $1.592\times10^9$ & $1.619\times10^{-3}$  \\ 
 $4.47\times10^6\;\mathrm{eV}$ & 1.745 & 1.877 & 14.47 & $67.73$ & $1.982\times10^8$  & $2.016\times10^{-4}$ \\ 
 $4.47\times10^4\;\mathrm{eV}$ & 1.741 & 1.872 & 14.47 & $6.639\times10^3$ & $1.966\times10^6$ & $2.000\times10^{-6}$ \\  $4.47\times10^2\;\mathrm{eV}$ & 1.741 & 1.872 & 14.47 & $6.637\times10^5$ & $1.965\times10^4$ & $1.999\times10^{-8}$ \\ 
 $4.47\;\mathrm{eV}$ & 1.741 & 1.872 & 14.47 & $6.637\times10^7$ & $1.965\times10^2$ & $1.999\times10^{-12}$ \\ 
 $4.47\times10^{-2}\;\mathrm{eV}$ & 1.741 & 1.872 & 14.47 & $6.637\times10^9$ & $1.965$ & $1.999\times10^{-12}$ \\  $\Lambda_\mathrm{DE}$ & 1.741 & 1.872 & 14.47 & $1.327\times10^{11}$ & $9.825\times10^{-2}$ & $9.994\times10^{-14}$ \\[0.5ex]
  \hline\hline
\end{tabular}
\captionsetup{width=.82\linewidth}
\caption{In this table, we are showing the mass at spatial infinity $\tilde{M}_\infty$, the baryon mass $\tilde{M}_b$, the stellar radius $\tilde{r}_*$, and the screening radius $\tilde{r}_k$ of seven different solutions for varying $\Lambda$. We also show the scalar field energy at spatial infinity normalized by either $\Lambda_\mathrm{DE}$ or $M_\mathrm{GR}$, in the Jordan frame ($\tilde{E}^\varphi_\infty$). The central density of the stars is fixed to $\rho_c=9.3\times10^{14}\;\gcc$, and the conformal coupling constant to $\alpha\approx0.14$.}
\label{scalarenergytable}
\end{center}
\end{table*}

\section{Non-linear evolution in spherical symmetry}
\label{dynamicsection}

In this section, we describe the formalism that we employ to perform fully non-linear numerical
evolutions in $k$-essence theory. We use as initial data  the static solutions presented in Sec.~\ref{sectionID}, subject to suitable initial perturbations
that trigger stellar oscillations or spherical collapse. We  present results for 
the evolution and show that gravitational collapse generically leads to diverging characteristic velocities, which can be avoided by adding a ``fixing equation'' in the spirit of the approach of Refs.~\cite{Cayuso:2017iqc,Allwright:2018rut}.

\subsection{Evolution formalism: spherical symmetry}

The covariant field equations \eqref{fieldeqs1}-\eqref{fieldeqs2} and   \eqref{fieldeqs4}-\eqref{fieldeqs3} can be written as an evolution system by splitting explicitly the spacetime into a foliation of space-like hypersurfaces with a normal time-like vector. Assuming spherical symmetry, we can adopt the line element 
\begin{equation}\label{metric_ansatz}
ds^2 = -N^2(t,r)dt^2+g_{rr}(t,r)dr^2+r^2g_{\theta\theta}(t,r)d\Omega^2 \,,
\end{equation}
where $N(t,r)$ is the lapse function, while $g_{rr}(t,r)$ and $g_{\theta\theta}(t,r)$ are positive metric functions. These quantities are defined
on each spatial slice with normal $n_{\mu}=(-N,0)$ and extrinsic curvature $K_{ij} \equiv -\frac{1}{2}\mathcal{L}_{n}\gamma_{ij}$, where $\mathcal{L}_{n}$ is the Lie derivative along $n^{\mu}$ and $\gamma_{ij}$ is the metric induced on each  spatial slice.

The Einstein equations \eqref{fieldeqs1} can be written as a hyperbolic evolution system by using the Z3 formulation~\cite{Bona:2002fq}, in which the momentum constraint is included in the evolution system by considering an additional vector $Z_{i}$ as an evolution field~\cite{Alic:2007ev,Bona:2005pp,Bernal:2009zy,ValdezAlvarado:2012xc}. Equation \eqref{fieldeqs1} can be expressed  as a first order system by introducing the following first derivatives of the fields as independent variables, 
\begin{eqnarray}
A_{r}=\frac{1}{N}\partial_{r}N \,,\quad  &&{D_{rr}}^{r}=\frac{g^{rr}}{2}\partial_{r}g_{rr} \,, \quad  {D_{r\theta}}^{\theta}=\frac{g^{\theta\theta}}{2}\partial_{r}g_{\theta\theta} \,,\nonumber \\
%%%%%%%%%%%%%%%%%%%%%
\chi&=&\partial_{r}\varphi \,, \qquad \Pi=-\frac{1}{N}\partial_{t}\varphi \,.
\end{eqnarray}
A coordinate system for the lapse (i.e. slicing condition) is required to close the evolution system. We use the singularity-avoidance $1+\log$ slicing condition $\partial_{t}\ln N=-2\,\mathrm{tr}K,$
where $\mathrm{tr}K=K_{r}^{r}+2K^{\theta}\,_{\theta}$~\cite{Bona:1994dr}. 
The final set of evolution fields for the Z3 formulation in spherical symmetry can be found in Ref.~\cite{ValdezAlvarado:2012xc}. 

The equation of motion~\eqref{fieldeqs2} for the scalar field becomes
\begin{eqnarray}
\partial_{t}\varphi &=& -N\Pi\label{KG2-1} \,,\\
%%%%%%%%%%%%%%%%%%%%%
\partial_{t}\chi &=& -\partial_{r}\left(N\Pi \right) \,,\\
%%%%%%%%%%%%%%%%%%%%%
\partial_{t}\Psi &=& -\partial_{r}F_{\Psi}^{r} -\frac{2}{r}F_{\Psi}^{r} + \frac{1}{2}N \zeta  \mathcal{A} T \,,\label{KG2-3}
\end{eqnarray}
\\
where $\zeta =\sqrt{g_{rr}}g_{\theta\theta}$ and
\begin{eqnarray}
\Psi&=&\zeta  K' \Pi\,, \label{fPi} \\
F_{\Psi}&=&N\zeta  K' g^{rr}\chi \,.
\end{eqnarray}
Note that we have introduced a new conserved field~$\Psi$, depending implicitly on the primitive fields $\{\Pi,\chi\}$ through the non-linear equation~\eqref{fPi}. In fact, during the evolution, this equation has to be solved numerically at each time-step to recover $\Pi$ (for further discussion see Ref.~\cite{Bezares:2020wkn}).

Finally, the conservation of the stress-energy tensor and of the baryon number, Eqs.~\eqref{fieldeqs4}-\eqref{fieldeqs3}, can be written as a (first-order) evolution system by splitting the four-velocity vector into its components parallel and orthogonal to the vector $n^{\mu}$, namely $u^{\mu} = W(n^{\mu} + v^{\mu})\,$,
being $W=-n_{\mu}u^{\mu}$ the Lorentz factor and $v^{\mu}$ the spatial velocity measured by  Eulerian observers. Assuming again spherical symmetry, the conservation equations  \eqref{fieldeqs4}-\eqref{fieldeqs3} become
\begin{widetext}
	\begin{eqnarray}
	\partial_{t}(\zeta  D) &=& -\partial_{r}(\zeta  D N v^r)+N\mathcal{A}\zeta  D(-\Pi+v^r \chi) - \frac{2}{r}\zeta  D N v^{r} \,,\\
%%%%%%%%%%%%%%%%%%%%%%%%%%%%%%%%%%%%%%%%%%%%%%%%%%%%%
	\partial_{t}(\zeta  U)
	&=& -\partial_{r}(\zeta  N {S}^{r}) + 
	N \zeta  \mathcal{A}   \Pi T +
	\zeta  N\left[ {S^{r}}_{r}{K^{r}}_{r}+2{S^{\theta}}_{\theta}{K^{\theta}}_{\theta}-{S}^r\left(A_{r} + \frac{2}{r}\right)\right] \,,\\
%%%%%%%%%%%%%%%%%%%%%%%%%%%%%%%%%%%%%%%%%%%%%%%%%%%%%%
	\partial_{t}(\zeta {S}_r)
	&=&-\partial_{r}(\zeta  N {S^r}_{r}) + N \zeta  \mathcal{A} \chi T  + \zeta N  \left[{{S}^{r}}_{r}\left({D_{rr}}^{r}-\frac{2}{r}\right)+2{S^{\theta}}_{\theta}\left({D_{r\theta}}^{\theta}+\frac{1}{r}\right)-U A_{r}\right]\,.
\end{eqnarray}
\end{widetext}

The evolved conserved quantities $\{\zeta D,\zeta U,\zeta S_{r}\}$ are respectively proportional to the rest-mass density measured  by  Eulerian  observers ($D$),  the  energy  density ($U$)  and  the  momentum density ($S_r$). These quantities, together with the non-trivial spatial components of the stress-energy tensor, can be written in terms of the physical (or primitive) fluid fields as
\begin{eqnarray}
D &=& \rho_{0} W\,,~{S}_r =  h W^2 v_r\,,~ U =  h W^2 - P\,,\label{con1}\\
{S_{r}}^{r} &=& hW^{2}v_{r}v^{r} + P\,,~{{S}_{\theta}}^{\theta} =  P\,,\label{con2}
\end{eqnarray}
where $h \equiv \rho_{0} (1 + \epsilon) + P$ is the enthalpy, $v^{r}$ is the radial velocity
and the Lorentz factor is simply $W^2= 1/(1-v_{r}v^{r})$. 

Note that,  during the evolution, one needs to recover the primitive fields $\{\rho_{0},\epsilon,P,v^{r}\}$ in order to calculate the right-hand-side of the evolution equations for the conserved fields $\{D,U,S_{r}\}$. This can only be achieved by including a closure relation between the pressure and the other thermodynamic fields. Here, we close the system by employing (both in the Jordan and the Einstein frame) the ideal fluid equation of state $P= (\Gamma -1)\rho_{0}\epsilon$, where $\Gamma$ is the same adiabatic index used for generating the initial data. 
Furthermore, as in the case of the scalar field,  the transformation from conserved to primitive fields requires to solve non-linear equations, which we do numerically at each time-step. For further discussion about the algorithm to convert  from conserved  to  primitive fields, we refer the interested reader to Ref.~\cite{ValdezAlvarado:2012xc}.

Finally, the complete evolution system is written in flux-conservative form
\begin{equation}
\partial_{t}{\bf u} + \partial_{r} F( {\bf u} ) = \mathcal{S}( { \bf u}) \,,
\end{equation}
where ${\bf u}
=\{N\,, g_{rr}\,, g_{\theta\theta}\,, {K_{r}}^{r}\,, {K_{\theta}}^{\theta}\,,A_{r}\,, {D_{rr}}^{r}\,, {D_{r\theta}}^{\theta}\,, Z_{r}, \\ \varphi\,,\Pi\,,\Psi\,,D\,,U\,,S_{r}\}$ is a vector containing the full set of evolution fields, and neither the radial fluxes $F( {\bf u} )$ nor the source terms $\mathcal{S}( {\bf u})$ contain terms with derivatives of the evolution fields.

\subsection{Numerical setup and radiation extraction}\label{numesetup}
The numerical code employed in this work is an extension of the one presented in Ref.~\cite{Bezares:2020wkn},
which was used to study the dynamics of  $k$-essence in vacuum spacetimes, with the model given by Eq.~\eqref{kessence}. The code has been fully tested also in GR, by studying the dynamics of black holes~\cite{Alic:2007ev}, boson stars~\cite{Bernal:2009zy}, fermion-boson stars~\cite{ValdezAlvarado:2012xc} and anisotropic compact objects~\cite{Raposo:2018rjn}.

We use a high-resolution shock-capturing (HRSC) scheme, based on finite-differences, to discretize both the Einstein  equations and the relativistic hydrodynamics equations~\cite{Alic:2007ev}. This method can be interpreted as a fourth-order finite difference scheme plus a third-order adaptive dissipation. The dissipation coefficient is given by the maximum propagation speed at each grid point. For the scalar field we use a more robust HRSC second-order method, by combining the Lax-Friedrichs flux formula  with a monotonic-centered limiter~\cite{CCC,Palenzuela:2018sly}.

The time evolution is performed through the   method of lines using a third-order accurate strong stability preserving Runge-Kutta integration scheme. We set a Courant factor $\Delta t/\Delta r = 0.125$, in units $G=c=M_{\odot}=1$, so that the CFL condition imposed by the principal part of the evolution system is always satisfied. Most of the simulations presented in this work have been performed with a spatial resolution of $\Delta r =0.008M_{\odot}$, in a domain with outer boundary located at $r=480M_{\odot}$. We use maximally dissipative boundary conditions for the spacetime variables, and outgoing boundary conditions for the scalar field.
We have verified that the results do not vary significantly when the position of the outer boundary is changed. 
We have also performed evolutions with different resolutions, which indicate that the results presented
here are consistent and within the convergent regime.

%%%%%%%%%%%%%%%%%%%

%\subsection{Scalar Gravitational Radiation}
Unlike in GR, monopole gravitational radiation (in the form of scalar field waves)
is permitted in ST  theories, and is produced 
by gravitational collapse in FJBD theories~\cite{Novak:1997hw,Gerosa:2016fri,Rosca-Mead:2020ehn}.
In the following we will see that a non-vanishing 
monopole flux is also emitted by stellar oscillations and by gravitational collapse   (in spherical symmetry) in $k$-essence.
The response of a gravitational interferometer to scalar waves 
is encoded in the Jordan-frame Newman-Penrose invariant 
$\phi_{22}$~\cite{PhysRevD.8.3308}, which far from the source can be
computed simply as~\cite{Barausse:2012da}:
\begin{eqnarray}\label{phi22_def}
\phi_{22} \simeq - \alpha \sqrt{16\pi\,G}\partial_t^2\varphi + O\left(\frac{1}{r^2}\right)\,.
\end{eqnarray}
In deriving this expression, Ref.~\cite{Barausse:2012da}
assumed a decay $\propto 1/r$ for the scalar field,
which, as stressed already, is only a good approximation
outside the screening radius in $k$-essence. For this reason,
and because the distance of the interferometer from the source
is typically much larger than the screening radius (even for $\Lambda \sim \Lambda_{\rm DE}$), we only compute
$\phi_{22}$ at extraction radii $r_{\rm ext}> r_k$. From 
$\phi_{22}$  one can then obtain the scalar strain $h_s$ via $\phi_{22}\propto\partial_t^2 h_s$ [which, by virtue
of Eq.~\eqref{phi22_def}, yields $h_s(r_{\rm ext})\propto \varphi(r_{\rm ext})$, up to 
terms constant and linear in time]. The scalar strain can in turn be used to compute the signal-to-noise ratio (SNR) for a given detector~\cite{Novak:1997hw,Gerosa:2016fri}.

%%%%%%%%%%%%%%%%%

\subsection{Stellar oscillations}
\label{stellaroscillations}

The non-linear stability of $k$-mouflage stars in equilibrium configurations, like those constructed in Sec.~\ref{sectionID}, can  be  tested by perturbing them and following their evolution numerically  using the formalism described above. 
Here, we  consider $k$-essence theories with  conformal coupling $\alpha\approx0.14$, but differing for the value of $\Lambda$, which we fix to
 either $\Lambda\sim 71.8$ MeV or  $\Lambda\sim 4.04$ MeV. The former 
gives rise to stars that are very similar to
 solutions of FJBD theory (with the same conformal coupling), while the latter produces a rather significant screening effect 
 on the scalar field (cf.  Sec.~\ref{sectionID}). Notice that we cannot consider $\Lambda$ as small as $\Lambda_{\rm DE}$,
 because, even though we can simulate static stars for this value of the strong-coupling scale, the corresponding
 dynamical evolutions become intractable because of large round-off errors (since, as already mentioned and detailed in Appendix \ref{app}, the hierarchy of scales between the screening and stellar radii 
 requires one to use code units $G=c=M_\odot=1$, in which $\Lambda_{\rm DE}\sim 10^{-12}$).
 Moreover, as shown in Ref.~\cite{terHaar:2020xxb}, 
 simulations of  stars with significant screening are
 also challenging as they require significant spatial resolution near the 
 origin, where
 the solutions pass from the 
 non-linear regime applicable to the outer layers of the star
 to a FJBD-like behavior.

\begin{figure*}
    \centering
    \includegraphics[height=0.5\textwidth]{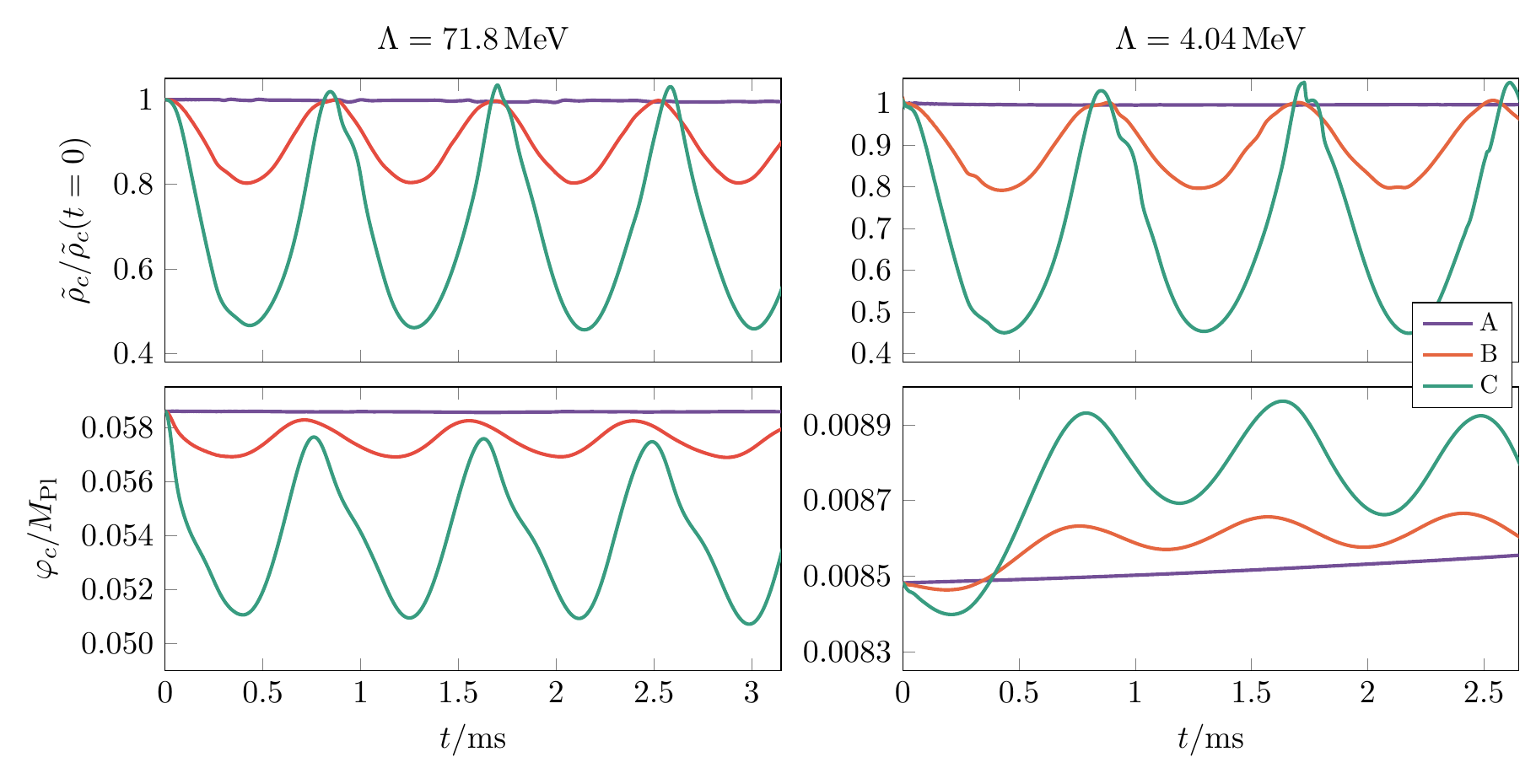} \caption{ Evolution of  the rest-mass density and the scalar field in the Jordan frame as a function of time for $\Lambda =71.8\,$MeV (left panel) and $\Lambda =4.04$\,MeV (right panel), and conformal coupling $\alpha\approx 0.14$. We consider static initial conditions (A), as well as small  (i.e. 4\%) and large (i.e. 14\%) initial perturbations in the internal energy density (B and C, respectively). Note that no secular growth is present, i.e.
$k$-mouflage stars are non-linearly stable.} 
\label{evol_c1e1020}
\end{figure*}

We consider equilibrium configurations with central energy density $\rho_c=9.3\times10^{14}\;\gcc$,
and excite oscillations by increasing  the internal energy by $4\,\%$ (``small oscillations'') or $14\,\%$ (``large oscillations''). Notice that although this initial perturbation  introduces small constraint violations, these are comparable to the solution's truncation error. Therefore, it is not  necessary to solve the energy constraint on the initial slice. Results for the two values of $\Lambda$ are presented in Fig.~\ref{evol_c1e1020}, which displays the central values for the rest-mass density and for the scalar field as a function of time.  The purple lines show the dynamics of unperturbed stars (i.e., stars only perturbed by numerical truncation errors), which confirms 
 the stability of these systems. For small perturbations  (red lines) and large perturbations (green lines),  the stars begin to oscillate. Indeed, 
since we increase the internal energy of the stars to  trigger the oscillations,
the stellar compactness initially decreases, and so does the scalar field magnitude. 
The latter  oscillates  with the same frequency as the density, but with a small time shift.
Notice that the oscillations do not grow in amplitude, confirming that these stars are stable.

\begin{figure}[h]
    \centering
    \includegraphics[height=0.4\textwidth]{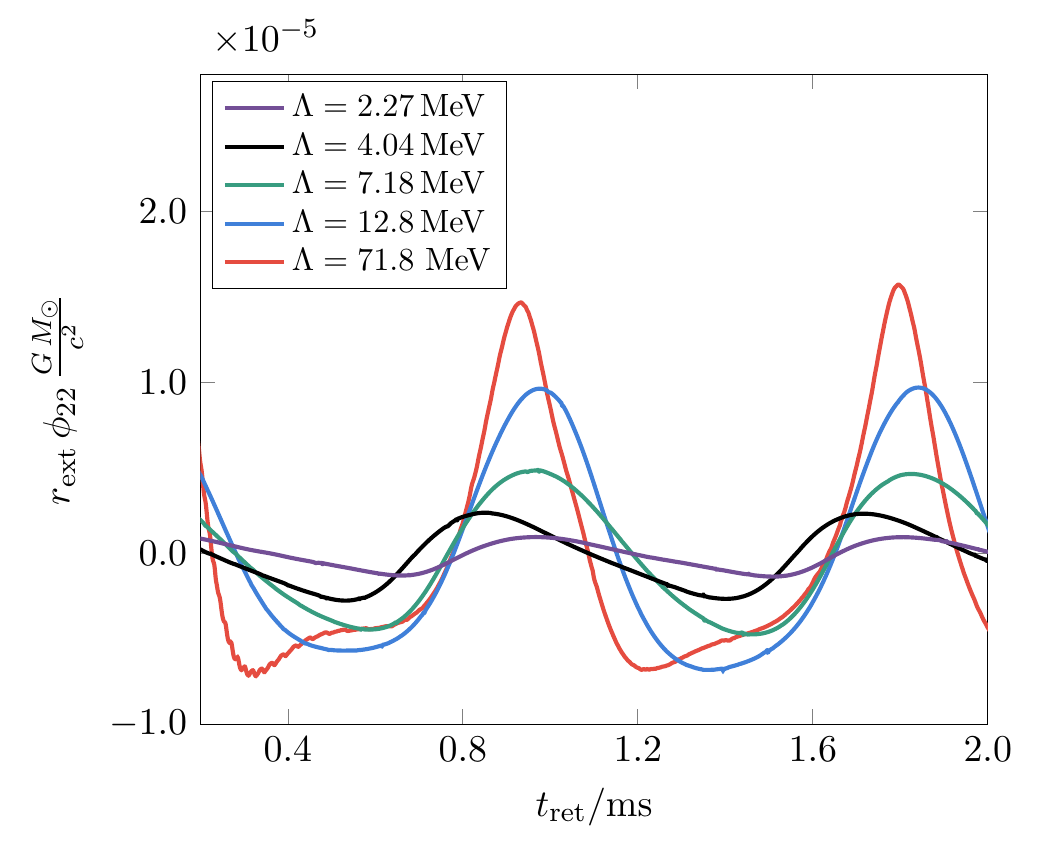}   \caption{The Jordan-frame Newman-Penrose invariant  $\phi_{22}$ (which describes monopole scalar radiation) for oscillating stars (with large 14\% initial perturbations in the internal energy density), as function of the retarded time  $t_{\rm ret}=t-r_{\rm ext}$, with $r_{\rm ext} =150\;GM_{\odot}> r_k$ the extraction radius. The conformal coupling is set to $\alpha\approx0.14$.}
    \label{radiation1}
\end{figure}

As can be seen from  Fig.~\ref{evol_c1e1020} (right panel), 
the amplitude of the central scalar field oscillations decreases with $\Lambda$, 
just like the central scalar field of the static solutions [cf. Eq. \eqref{relphi}]. This  seems to confirm the validity of kinetic screening even in this dynamical case. 
To strengthen this conclusion, we have also extracted the scalar monopole signal $\phi_{22}$
for oscillating stars initially subjected to the same large ($\sim 14$\% ) perturbations 
of the internal density, for $\Lambda=\{71.8,12.8,7.18,4.04,2.27\}$ MeV. The results are
presented in Fig.~\ref{radiation1} for an extraction radius $r_{\rm ext}=150 G M_\odot> r_k$, as
a function of retarded time, defined as $t_{\rm ret}=t-r_{\rm ext}$. As can be seen, the amplitude of the signal
is an increasing function of $\Lambda$. 

\begin{figure*}
    \centering
    \includegraphics[height=0.4\textwidth]{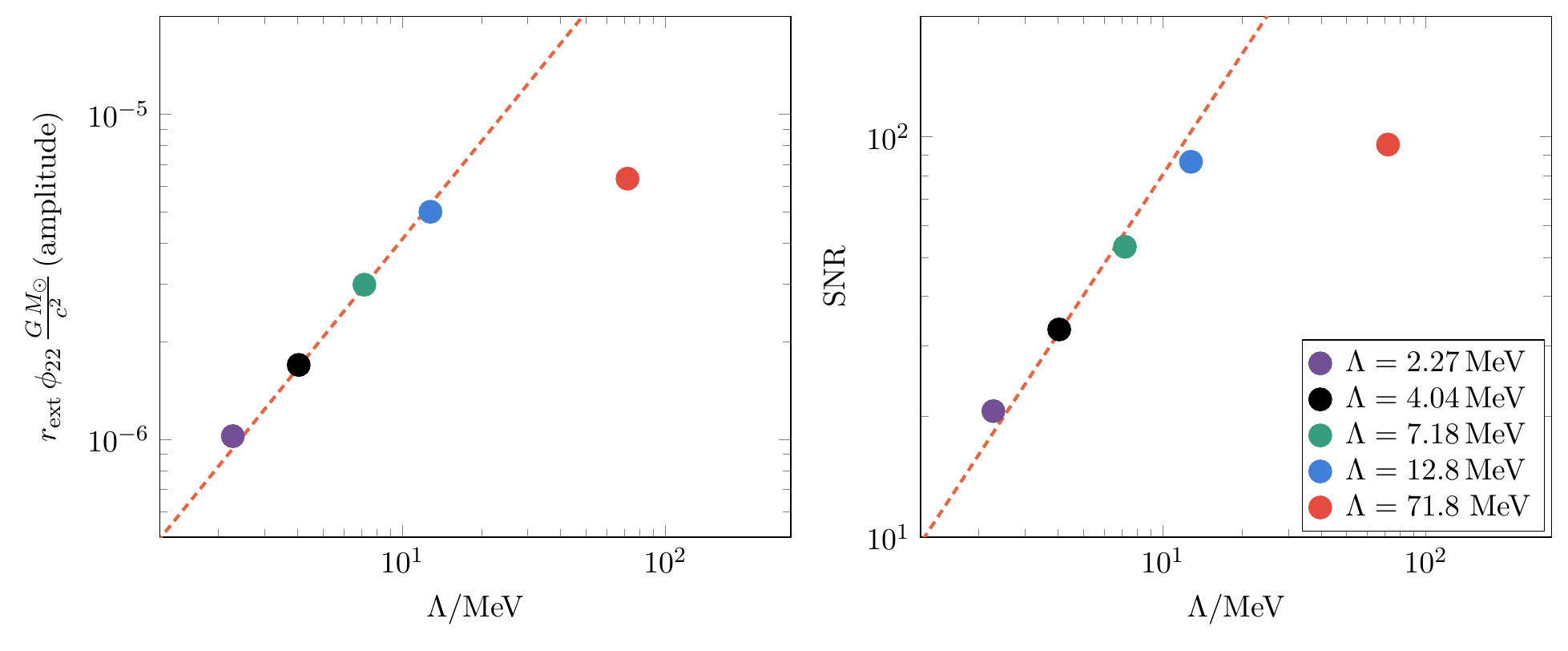}   \caption{ The amplitude of $r_{\rm ext}\,\phi_{22}$ as plotted in Fig.~\ref{radiation1}
    (left panel) and the corresponding SNR at 8 kpc for Advanced LIGO at design sensitivity (right panel) as a function $\Lambda$. The orange dashed lines show a linear scaling in $\Lambda$.}
    \label{radiation2}
\end{figure*}

In order to see the effect of screening more clearly,
we have plotted in Fig.~\ref{radiation2} (left panel) the amplitude of the same signals, which
we compute as the root mean square of the time series. Notice that with the exception of
$\Lambda\sim 71.8$ MeV, for which there is no screening (even in the static case), the monopole amplitude scales as $\Lambda$, as expected from the scaling of the central scalar field of the static stellar solutions [cf. Eq. \eqref{relphi}].
 By integrating $\phi_{22}$ in time twice to get the monopole strain $h_s$, we can compute its SNR for Advanced LIGO (at design sensitivity\footnote{For the sensitivity, we used the zero detuning, high power configuration of \url{https://dcc.ligo.org/LIGO-T0900288/public}.}) for an optimally oriented source at 8 kpc  (corresponding to the distance between the Earth and the center of the Galaxy).
 The results are displayed  in Fig.~\ref{radiation2} (right panel)
 and show again a scaling roughly linear with $\Lambda$.
 Extrapolating to values of $\Lambda\sim \Lambda_{\rm DE}$ relevant for dark energy,
 one would get a tiny unobservable SNR $\sim 10^{-6}$ at 8 kpc.

%%%%%%%%%%%%%%%%%%%%%%%%%%%%%%%

\subsection{Gravitational collapse}

As  discussed in Ref.~\cite{terHaar:2020xxb},  the characteristic propagation speeds of the scalar field equation~\eqref{fieldeqs2} diverge when $k$-mouflage stars collapse (``Keldysh problem''). In more detail, the evolution equation for the scalar field can be recast as 
\begin{eqnarray}
\gamma^{\mu\nu}\nabla_\mu \nabla_\nu \varphi &=& \frac{\mathcal{A}\,T}{2\,K'(X)} \,, \label{eqKG}
\end{eqnarray}
in terms of the effective metric 
\begin{equation}
    \gamma^{\mu\nu} \equiv g^{\mu\nu} + \frac{2 K''(X)}{K'(X)}{\de}^{\mu}\varphi\de^{\nu}\varphi\,.
\end{equation} 
The characteristic speeds of this equation are then given by~\cite{Bezares:2020wkn} 
\begin{eqnarray}
V_\pm = -\frac{ \gamma^{tr}}{ \gamma^{tt}}\pm \sqrt{\frac{-{\rm det}( \gamma^{\mu\nu})}{( \gamma^{tt})^2}} \,.
\end{eqnarray}
As shown in Ref.~\cite{Bezares:2020wkn}, at leading order (on
Minkowski space and in standard Cartesian coordinates)
these velocities reduce to the usual expression for the 
speed of the scalar mode in $k$-essence, $c_s=\pm\sqrt{1+2XK''/K'}$ (see e.g. \cite{Babichev:2007dw}), of which they constitute the non-linear generalization.

During the gravitational collapse of a $k$-mouflage star (which can be triggered e.g. by decreasing its internal energy), these velocities diverge because $\gamma^{tt}$ goes to zero. This problem  also appears during the collapse of scalar field pulses in vacuum~\cite{Bezares:2020wkn,Bernard:2019fjb,Figueras:2020dzx}, and resembles the
behavior of the Keldysh equation
\begin{equation}
t\,\partial^2_t\varphi(t,r) + \partial^2_r\varphi(t,r) = 0 \,.
\end{equation}
This equation is hyperbolic with characteristic speeds $\pm(-t)^{-1/2}$ for $t<0$, leading
to a divergence at $t=0$.

Diverging characteristic speeds constitute at the very least a practical obstacle that
prevents one from evolving the dynamics past this divergence by using explicit time integrators, since the CFL bound
forces the time step to vanish when the Keldysh behavior appears.
As stressed in Ref.~\cite{Bezares:2020wkn}, this divergence may in principle be avoided by 
allowing for a non-vanishing shift. However, neither Ref.~\cite{Bezares:2020wkn} nor Ref.~\cite{terHaar:2020xxb} managed to find a  suitable coordinate condition in spherical symmetry that
would maintain the characteristic speeds  finite while still ensuring stable numerical evolutions. This leaves open the possibility that the Keldysh problem that we find might have a physical relevance, besides a practical one.
Here, however, we assume that the Keldysh problem is not fundamental, and we attempt to 
amend it by using 
an approach inspired by Refs.~\cite{Cayuso:2017iqc,Allwright:2018rut},
which put forward a method to ameliorate the stability of Cauchy evolutions in theories with higher derivatives\footnote{This method is in turn inspired  by  the M\"{u}ller–Israel–Stewart formalism of viscous relativistic hydrodynamics~\cite{Muller:1967zza,Israel:1976tn,1976PhLA...58..213I}.} (see also Ref.~\cite{Cayuso:2020lca} for an application of this approach to a specific higher derivative extension of GR). 

The method consists of modifying the theory's dynamics by adding extra fields and 
``fixing equations'' (i.e. drivers) for them. The drivers are chosen so that on sufficiently long timescales the evolution dynamics approximately matches that of the theory under consideration ($k$-essence in our case).
We stress that this modification of the field equations does not correspond to a standard ultraviolet completion of $k$-essence, which is not known for theories giving screening~\cite{Adams:2006sv}. 
However, this dynamical fixing of the Cauchy problem 
might make sense if the  effective field theory 
 ``classicalizes''~\cite{Dvali:2016ovn}  at high energies.

To apply the method of Refs.~\cite{Cayuso:2017iqc,Allwright:2018rut}, 
let us first recall that  Ref.~\cite{terHaar:2020xxb} found that
in order to deal with shocks appearing in $k$-mouflage stars, the scalar field equation needs to be written as a conservation law [cf. Eq.~\eqref{fieldeqs2}]. The ``fixing equation'' that we introduce must therefore share this property. 
Let us then introduce the new field $\Sigma$ and the modified evolution system
\begin{align}
&\partial_t \left(\sqrt{-g} \Sigma \nabla^t\varphi \right) + \partial_i \left( \sqrt{-g} \Sigma \nabla^i\varphi \right) = \frac{1}{2}\sqrt{-g}\mathcal{A}T \,,
\label{eq8}\\
&\partial_t \Sigma = - \frac{1}{\tau}  (\Sigma - K'(X)) \,. \label{eq9}
\end{align}
 The second equation is a driver that will force  $\Sigma$ to $K'(X)$ on a timescale $\tau>0$. As can be seen, the principal part of this system takes indeed the form of a conservation law.
Restricting then to the spherical symmetric case and using the line element~\eqref{metric_ansatz}, Eqs.~\eqref{eq8}-\eqref{eq9} can be written as
\begin{eqnarray}
	\partial_{t}\varphi &=& -N\Pi\label{MIS1} \,,\\
	%%%%%%%%%%%%%%%%%%%%%%%%%%%%%%%%%%%%%%%%%%%%%%%%%%%%%%%%%%%%%%%%%%%%%%%%%%
	%%%%%%%%%%%%%%%%%%%%%%%%%%%%%%%%%%%%%%%%%%%%%%%%%%%%%%%%%%%%%%%%%%%%%%%%%%
	\partial_{t}\chi &=& -\partial_{r}\left[N\Pi \right] \,,\\
	%%%%%%%%%%%%%%%%%%%%%%%%%%%%%%%%%%%%%%%%%%%%%%%%%%%%%%%%%%%%%%%%%%%%%%%%%%
	%%%%%%%%%%%%%%%%%%%%%%%%%%%%%%%%%%%%%%%%%%%%%%%%%%%%%%%%%%%%%%%%%%%%%%%%%%
	\partial_{t}\Psi &=& -\partial_{r}F_{\Psi}^{r} -\frac{2}{r}F_{\Psi}^{r} + \frac{1}{2}N \zeta  \mathcal{A} T \,, \\
	\partial_t \Sigma &=& - \frac{1}{\tau}  \left(\Sigma - K'(X) \right) \,,\label{MIS4}
\end{eqnarray}
where $\Psi = \zeta\,\Sigma\,\Pi$ and $F_{\Psi}^{r} = N\,\zeta\,\Sigma\,g^{rr}\,\chi\,$. As in the original $k$-essence equations in balance law form~\cite{Bezares:2020wkn}, there is a set of conserved evolved fields $\{\chi,\Psi, \Sigma\}$ and a set of primitive fields $\{\chi,\Pi,\Sigma\}$ required to calculate the right-hand side of the equations. In this case, the only unknown primitive field ($\Pi$) can be found by solving the  linear equation $\Pi =  \Psi/\zeta \,\Sigma$ at each time-step. Finally, notice that the evolution equations~\eqref{MIS1}-\eqref{MIS4} lead to a strongly hyperbolic system,
thus ensuring that the Cauchy problem is well-posed. 
We stress that this approach works trivially for FJBD theories, since for the latter
$K'(X)=-1/2$ is constant, and the driver Eq.~\eqref{MIS4} relaxes $\Sigma$ to $K'(X)$ exponentially on the timescale $\tau$.
Moreover, we have tested it against the oscillating stars presented in the previous section obtaining very good agreement.

\begin{figure}[h]
    \centering
    \includegraphics[height=0.45\textwidth]{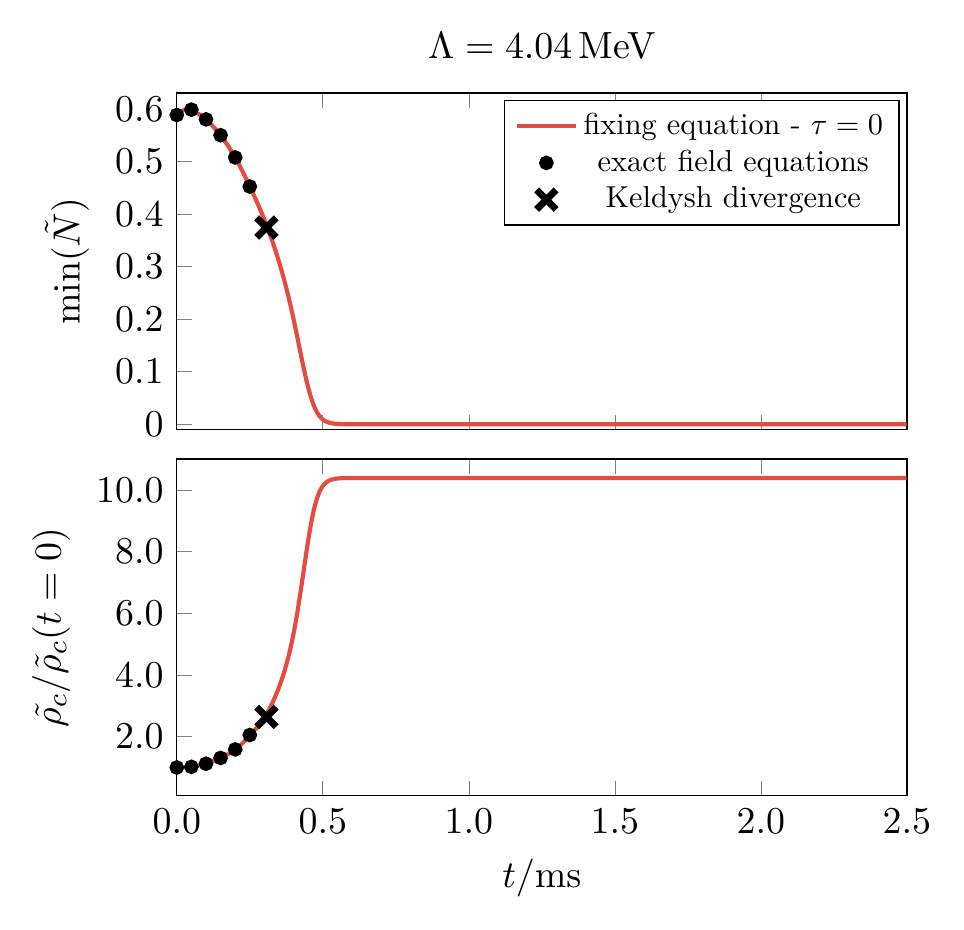}
    \caption{Evolution of the minimum of the lapse across the radial grid (top panel) and the central rest-mast density (bottom panel) in the Jordan frame, for the gravitational collapse of a neutron star in a theory with $\Lambda=4.04$\, MeV and $\alpha\approx0.14$. The red lines represent the evolution obtained with the fixing equation (extrapolated to $\tau=0$), and the  black circles  represent results obtained by solving the field equations \eqref{fieldeqs1}-\eqref{fieldeqs2}. Note 
    that the latter evolution presents diverging characteristic speeds for the scalar field at  $t=0.37$\,ms (``Keldysh behavior'', black cross), which effectively halts the simulation.}
    \label{collapsec1e20}
\end{figure}

Results for gravitational collapse in a theory with $\Lambda=4.04$ MeV are shown in Fig.~\ref{collapsec1e20},
for the minimum of the lapse (top panel) and the central rest-mast density (bottom panel).  The black circles  represent results obtained by solving the field equations \eqref{fieldeqs1}-\eqref{fieldeqs2}. In this case, the characteristic speeds of the scalar field diverge (at the time marked by a black cross) and the simulation stops long before formation of a horizon because of the CFL condition.  The solid red line shows instead the results obtained by adding the fixing equation,  which allows for the simulation to successfully complete, leading to the formation  of a hairless Schwarzschild BH. The results are obtained 
    for values of $\tau$ down to $30\,GM_\odot$,
    and are extrapolated to $\tau=0$.

\begin{figure}
    \centering
    \includegraphics[height=0.42\textwidth]{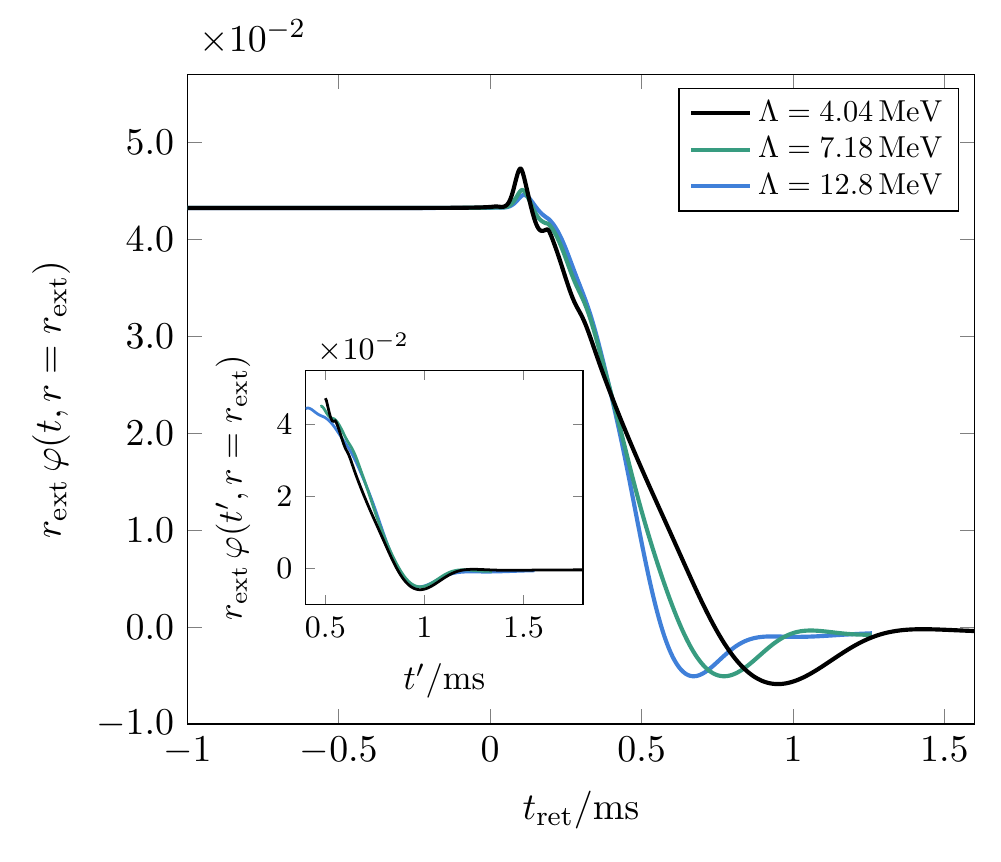}   \caption{Evolution of the scalar field far from the source as a function of the retarded time $t_{\rm ret}$ in the Jordan frame for different values of $\Lambda$ and $\alpha\approx0.14$. These results have been obtained by extrapolating to $\tau=0$. In the inset we display the scalar field as a function of the rescaled time $t'$, to show the self-similarity of these solutions during the gravitational collapse.}
    \label{new}
\end{figure}

Fig.~\ref{new} shows instead the time evolution of the scalar field far from the source (at an extraction radius $r_{ \rm ext}=200\,G M_{\odot}>r_k$)  as a function of time, for three values of $\Lambda$ giving screening in the static case     ($\Lambda=12.8,\,7.18,\,4.04$ MeV).
    The results are again obtained 
    for finite values of $\tau$ (as small as $10$ or $30$ $GM_\odot$ according to the value of $\Lambda$)
    and then extrapolated to $\tau=0$. As indicated, the scalar field  is multiplied by the extraction radius so that the value displayed is independent of
    the exact extraction position, i.e. we show $\varphi\, r_{\rm ext}$, with $r_{\rm ext}=200\,GM_\odot$.
    As can be seen, $\varphi\, r_{\rm ext}$ goes from a constant non-vanishing value
    at the beginning of the simulation to zero at late times, for all values of $\Lambda$. This behavior is readily explained. The initial value is set by the coefficient $\varphi_1$ of Eq.~\eqref{falloff}, which
    is proportional to the scalar charge [cf. Eq.~\eqref{chargedef}] and which is largely independent
of $\Lambda$, since scalar effects are not screened for $r>r_k$. The final value is zero
because a black hole forms, and in $k$-essence black holes have no hair (i.e. no scalar charge) because the theory is shift symmetric~\cite{Hui:2012qt,Sotiriou:2013qea}. Therefore, we can interpret the difference between the initial and final values of $\varphi r_{\rm ext}$ as due to the collapsing star shedding its scalar hair.

\begin{figure*}
    \centering
    \includegraphics[height=0.4\textwidth]{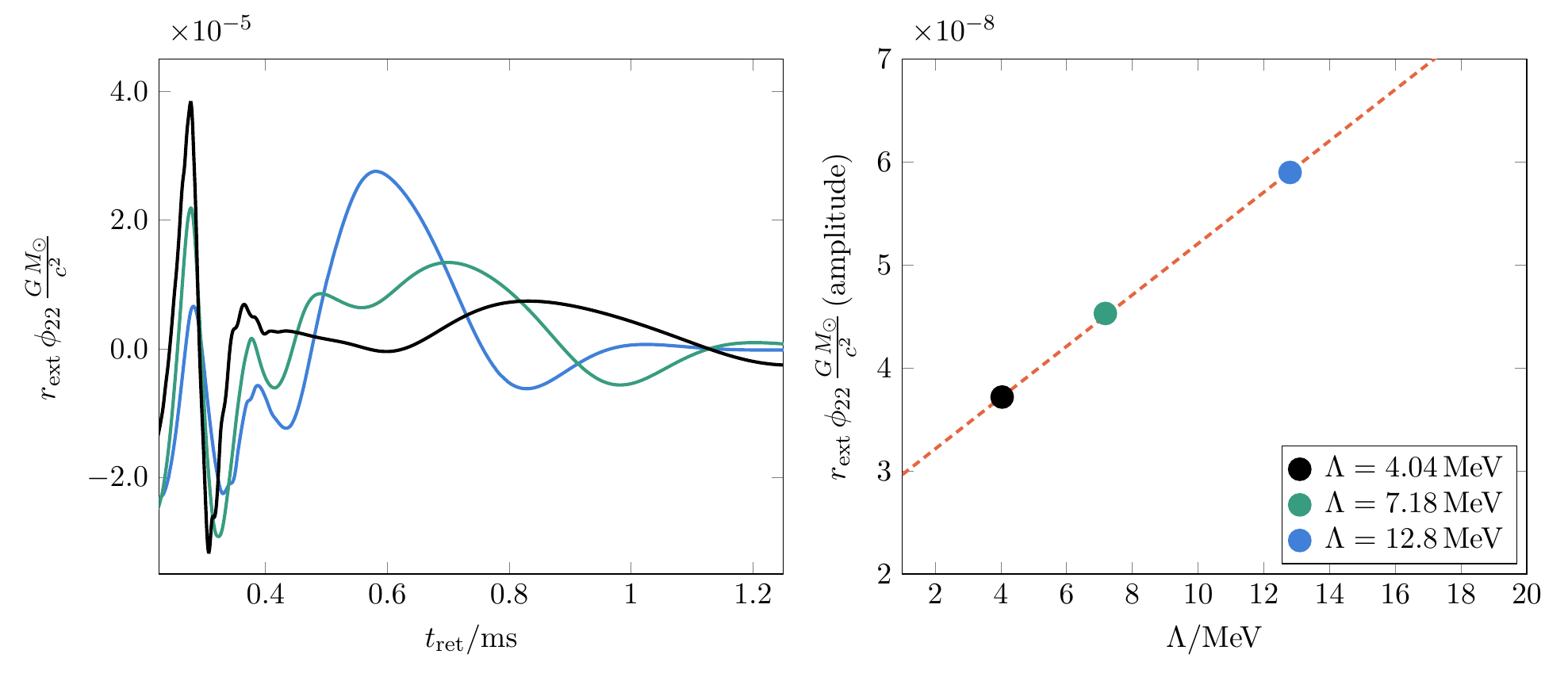} \caption{In the left panel, we show the Jordan-frame Newman-Penrose invariant  $\phi_{22}$ for collapsing stars, as function of the retarded time $t_{\rm ret}$, with $r_{\rm ext} =200\;GM_{\odot}> r_k$ and conformal coupling $\alpha\approx0.14$. On the right, we show the amplitude of $\phi_{22}$ as a function of $\Lambda$, together with a linear fit in $\Lambda$ (orange dashed line). 
    }
    \label{rms2}
\end{figure*}

Moreover, smaller values of $\Lambda$ seem to lead to longer characteristic timescales (i.e. lower frequencies) in the simulations of Fig.~\ref{new}. In fact, if one
plots the scalar field's evolution as function of a rescaled time $t'=(t-t_0)\sqrt{\Lambda} G^{1/4}$
(with $t_0$ a suitable offset), the results are very similar, as shown in the inset of
Fig.~\ref{new}. From this ``self-similarity'', we can conclude that
the frequencies contained in the signal should scale as $f\propto\sqrt{\Lambda}$. 
By combining this with the observation that the initial and final values of $\varphi$
are independent of $\Lambda$, we can infer that $\phi_{22}$ should scale with $\Lambda$ as $\phi_{22} \propto (2 \pi f)^2\varphi\propto\Lambda$. We have verified this scaling by computing $\phi_{22}$ explicitly (Fig.~\ref{rms2}, left panel), extracting its amplitude as the root mean square of its time series, and verifying that the amplitude scales roughly linearly with $\Lambda$ (Fig.~\ref{rms2}, right panel).
%%%%%%%%%%%%%%%%%%%%%%%%%%%%%%%%

As for the SNR of the results shown in Fig.~\ref{new}, 
we have computed it (assuming optimal source orientation) for Advanced LIGO at design sensitivity, and  obtained
 values of $\sim200$
 at  $8$ kpc, with no appreciable dependence on the value of $\Lambda$.
 This roughly constant (and detectable) SNR comes about because
 the difference between the initial and final value of $\varphi$ (and thus the scalar strain $h_s$) are largely independent of $\Lambda$, since the star has to shed all of its hair before forming a back hole. Because of the scaling of the frequency with $\sqrt{\Lambda}$, however, we expect that for  $\Lambda\to\Lambda_{\rm DE}$
 the signal will eventually fall out of the frequency band of terrestrial detectors. The latter are insensitive to frequencies lower than 1-10 Hz because of seismic noise
 (even for third generation detectors such as the Einstein Telescope~\cite{Punturo:2010zz} or Cosmic Explorer~\cite{Reitze:2019iox}). In fact, when going from $\Lambda\sim 10$ MeV for the results in Fig.~\ref{new}
 (whose frequencies are $\sim $ kHz) to $\Lambda\sim 10$ eV, we expect the frequency
 to drop by a factor $\sim 1000$ to $\sim 1$ Hz. Scalar monopole signals in theories with $\Lambda\lesssim 10$ eV
 are therefore likely  unobservable from Earth, but would fall in principle in the band of space-borne detectors such as LISA. By using the self-similarity of our solutions to compute the SNR for LISA in the case of $\Lambda\approx \Lambda_{\rm DE}\approx 2\,$meV, we
 obtain SNR$\,\sim 30$--40 (according to whether we use the LISA sensitivity curve  from the proposal to ESA~\cite{Audley:2017drz} or from the Science Requirements Document~\cite{scird}) for optimally oriented sources at 8 kpc distance. 
 For $\Lambda\approx 10\,$ meV, we get instead SNR$\,\sim 7$--10.
 We should stress again, however, that these results involve an extrapolation over nine orders of magnitude in $\Lambda$, based on the self-similarity of our simulations.

\section{Conclusions}
\label{conclusions}
In this work, we have studied the spherically symmetric non-linear dynamics of compact stars in ST theories with first-order derivative self-interactions for the scalar field ($k$-essence theories). These theories have been suggested to possess a mechanism ($k$-mouflage, or kinetic screening) that suppresses
the scalar fifth force on local (solar system) scales, while allowing for potentially significant scalar effects on large (cosmological) scales. We have confirmed that
$k$-mouflage works  for static spherically symmetric compact stars, whose structure we have calculated
exactly (up to numerical errors) for cosmologically relevant values ($\sim $ meV) of the theory's strong-coupling scale $\Lambda$.
These solutions are far from trivial to derive, because of the hierarchy of scales between the stellar radius and the screening one ($\sim 10^{11}$ km), but they confirm that no observable deviation from the GR geometry is to be expected
in the exterior of static spherically symmetric stars (whatever their compactness),
as long as one remains
within the screening radius.

We have then used these static spherically symmetric solutions as initial data for dynamical evolutions (again in spherical symmetry). In more detail, we have
triggered (non-linear) oscillations of our compact stars by perturbing their internal energy, and extracted the resulting monopole scalar radiation outside the screening radius. While we could not simulate theories with strong-coupling scales relevant for dark energy, 
we have managed to evolve stars in theories with $\Lambda$ as small as a few MeV, which already shows that kinetic screening suppresses the monopole scalar emission from stellar oscillations. Extrapolating to $\Lambda\sim$ meV, we have concluded that no observable monopole emission is to be expected from stellar oscillations in these theories.

We have also used our static spherically symmetric solutions as initial data for gravitational collapse. As reported in Ref.~\cite{terHaar:2020xxb},  the $k$-essence equations are always strongly hyperbolic, irrespective of the local state of the dynamical variables (at least if terms cubic in the scalar kinetic term are included in the action), but the characteristic speeds for the scalar field diverge during collapse. The same behavior appears in vacuum, for configurations close to critical collapse~\cite{Bezares:2020wkn,Bernard:2019fjb}. This divergence is at the very least a practical problem, as the system cannot be simulated past it because  of the CFL bound (i.e. the theory becomes non-predictive). While Ref.~\cite{Bezares:2020wkn} showed that 
the characteristic speeds can be maintained finite by allowing for a non-vanishing shift vector, it could not find a shift choice in spherical symmetry yielding stable evolutions. 

We have taken here a different approach, and modified the $k$-essence dynamics by introducing an auxiliary variable and a driver (or ``fixing equation'') that relaxes the modified dynamics to the true one on long timescales. We have done so in the spirit of the recent proposal by Refs.~\cite{Cayuso:2017iqc,Allwright:2018rut}, which is in turn inspired by dissipative relativistic hydrodynamics. This method has allowed us to simulate  gravitational collapse without incurring in any divergent characteristic speed, for strong-coupling scales as low as a few MeV. 
We have found that, unlike in the 
case of stellar oscillations, kinetic screening does \textit{not} suppress the monopole scalar radiation (extracted outside the screening radius) from the collapse. This happens because the collapsing star
must shed away all of its scalar hair in scalar waves before forming a (hairless) black hole.
This scalar signal would not be detectable 
by terrestrial gravitational wave detectors because its very low frequency (at least for values of $\Lambda\sim$ meV relevant for dark energy), but we
conjecture that it might be observable with space-based detectors such as LISA, if a supernova explodes in the Galaxy.

\appendix
\section{Units}
\label{app}

In this paper, we have used units $\hbar=c=1$, in which the $k$-essence action 
is given by Eq.~\eqref{action}. When simulating neutron stars numerically, it is convenient
to use units adapted to the problem, e.g. $G=c=M_\odot=1$. 

To see what the $k$-essence action is in these units, let us first factor out the Planck mass in the $k$-essence Lagrangian density:
\begin{equation}\label{actionapp}
    {\cal L}_k=\frac{1}{16 \pi G}\left(R 
    -\frac{1}{2}\bar{X}+\frac{\beta}{4\bar{\Lambda}^4}\bar{X}^2-\frac{\gamma}{8\bar{\Lambda}^8}\bar{X}^3+\ldots  \right)  \,,
\end{equation}
where we have introduced $\bar{X}\equiv 2 X/M_{\mathrm{Pl}}^2$, which is the kinetic energy
$\bar{X}\equiv {g}^{\mu\nu} \partial_\mu\bar{\varphi} \partial_\nu\bar{\varphi}$
for the dimensionless scalar $\bar{\varphi}\equiv \sqrt{2} \varphi/M_{\mathrm{Pl}} $, and defined
also $\bar{\Lambda}\equiv2^{1/4}\Lambda/{M_{\mathrm{Pl}}}^{1/2}$.

To reinstate $\hbar$, one can then note that in generic units the first two terms (the Ricci curvature and the kinetic energy for the rescaled dimensionless field) have dimensions of a length${}^{-2}$, hence one needs $\bar{\Lambda}=2^{1/4}\Lambda/(M_{\mathrm{Pl}}\hbar)^{1/2}$,
which has the correct dimensions of length${}^{-1/2}$ (with $c=1$).
For $\Lambda\approx \Lambda_{\rm DE} \sim 2 \times 10^{-3}\;\mathrm{eV}$,
one then has  $\bar{\Lambda}\sim 10^{-13} {\rm m}^{-1/2}$. In units $G=c=M_\odot=1$, lengths are measured in units of the Sun's Schwarzschild radius $G M_\odot/c^2\approx 1.5$ km, and therefore in these units one has 
 $\bar{\Lambda}\sim 4\times 10^{-12}$. Rewriting then the action \eqref{actionapp}
 in the same form as  Eq.~\eqref{action}, but in units $G=c=M_\odot=1$, one gets
 \begin{equation}\label{actionapp}
    {\cal L}_k=\frac{1}{16 \pi}R 
    -\frac{1}{2}{X}+\frac{\beta}{4{\Lambda}^4}{X}^2-\frac{\gamma}{8{\Lambda}^8}{X}^3+\ldots   \,,
\end{equation}
 where ${X}= \bar{X}/(16 \pi)$, ${\varphi}= \bar{\varphi}/\sqrt{16 \pi}$
 and ${\Lambda}= \bar{\Lambda}/(16 \pi)^{1/4}\approx 10^{-12}$.
  This very small value is among the reasons why numerical evolutions of the dynamics of collapsing or oscillating stars are challenging for theories with $\Lambda\sim \Lambda_{\rm DE}$. We stress, however, that we could successfully simulate
 static stars for such theories (thanks to Mathematica's~\cite{Mathematica} arbitrary machine precision arithmetic).

\begin{acknowledgments}
We thank L. Lehner for enlightening conversations on the well-posedness of the Cauchy problem.
M.B, L.t.H, M.C, and E.B. acknowledge support from the European Union's H2020 ERC Consolidator Grant ``GRavity from Astrophysical to Microscopic Scales'' (Grant No.  GRAMS-815673). 
C.P. acknowledges support from the Spanish Ministry of Economy and Competitiveness Grants No. AYA2016-80289-P and No. PID2019-110301GB-I00 (AEI/FEDER, UE).
M.B. acknowledges the support of the PHAROS COST Action (CA16214).
\end{acknowledgments}

\bibliographystyle{apsrev4-2}
\bibliography{master}

\end{document}